\documentclass[prb,twocolumn,amssymb,amsmath,superscriptaddress]{revtex4-1}
\usepackage{graphicx}
\usepackage{dcolumn}
\usepackage{bm}
\usepackage{color} 
\usepackage{braket}

\newcommand{\qvec}{{\bf q}}

\begin{document}
\title{Adiabatic transition from a BCS superconductor to a Fermi liquid and phase dynamics} 

\author{G.~Seibold} 
\affiliation{Institut F\"ur Physik, BTU Cottbus, PBox 101344, 03013 Cottbus,
Germany}
\author{C.~Castellani}
\affiliation{ISC-CNR and Department of Physics, University of Rome ``La
  Sapienza'',\\ Piazzale Aldo Moro 5, 00185, Rome, Italy}
\author{J.~Lorenzana}
\affiliation{ISC-CNR and Department of Physics, University of Rome ``La
  Sapienza'',\\ Piazzale Aldo Moro 5, 00185, Rome, Italy}

 
\begin{abstract}
  We investigate the physics of an adiabatic transition from a BCS
  superconductor to a Fermi liquid for an exponentially slow decreasing pairing
  interaction. In particular, we show that the metal keeps memory of the parent BCS state so it is possible to reverse the dynamics and go back to the original state similarly to a spin/photon echo experiment. Moreover, we study the
  evolution of the order  parameter phase $\phi$ in transforming the BCS superconductor to a
  conventional metal.  Since the global phase is the conjugate variable of the density we explicitly  show how to
  use the dynamics of $\phi$ together with gauge invariance to build up the
  non-interacting chemical potential away from particle-hole symmetry. 
  We further analyze the role of $\phi$ in restoring the gauge invariant
  current response when the non-interacting Fermi liquid is approached
  starting from a BCS superconductor in the presence of an external vector
  field.
\end{abstract}  

\maketitle

The quantum adiabatic theorem was proposed by Born and Fock \cite{born28} in 1928 and
sets out the basic principles for the evolution of the wave function
for a time dependent hamiltonian $H(t)$. Starting from the ground state
at $t = 0$ it states that the wave function passes the corresponding
ground states for all times providing that the change of $H(t)$ is
infinitely slow and there is no energy level crossing of ground states.

The adiabatic theorem is also at the heart of perturbative approaches in
quantum field many-body theory which rely on the Gell-Mann and Low theorem
\cite{gml51}, i.e. the evolution of the ground state $|\Phi_0\rangle$ of a
non-interacting system to an eigenstate $|\Psi\rangle$ of the
interacting system when an interaction parameter $V(t)=e^{-\varepsilon |t|} V(t=0)$ is turned on infinitely slowly, i.e. in the limit $\varepsilon\to 0^+$.
It is important to note that $|\Psi\rangle$ not necessarily corresponds
to the ground state $|\Psi_0\rangle$ of the interacting system,
in particular when the latter cannot be represented by a
perturbation series in the coupling constant. This is relevant 
in case of a superconductor, which is our focus, and  
where the order parameter $\Delta \sim e^{-1/\lambda}$ does not have a Taylor series in the coupling constant $\lambda\propto V$.

The rapid progress during the past two decades in the study
of ultracold Fermi gases \cite{giorg08,toermae18} has also stimulated
investigations of the adiabatic dynamics in the vicinity of quantum phase
transitions \cite{polkov05,dziarmaga05,zurek05} which are
usually associated with the vanishing of a gap so that the adiabaticity
condition cannot be satisfied. 
However, for a time dependent quench 
the number of excited states per unit volume vanishes as a power
law in the rate of change of the interaction parameter $V(t)$, so that for a 'sufficiently' slow
dynamics the system still follows an adiabatic process.

The experiments mentioned above have motivated the analysis of the BCS
pairing problem with time-dependent interactions and several 
proposals based on the realization of a suitable
out-of equilibrium dynamics (pump) which is then measured by
a probe pulse.~\cite{kuhn1,kuhn2,mazza12,manske14,Bunemann2017,mazza17,Collado2018,Collado2019,Collado2020,benfatto19,goetz20}
Within the pseudospin formulation
of Anderson \cite{and2} the problem can be mapped onto an effective
spin Hamiltonian for which the Bloch dynamics can be solved exactly.\cite{bara04,yus05,bara06,alt06,yus06} 

In the present paper we investigate the case of adiabatically switching off
the interaction in a BCS superconductor. Starting from the superconducting
ground state we are interested in how the Fermi liquid is approached as a
function of the adiabatic time scale. Indeed, within the context of the
equations of motion, the integrability of the BCS model leads to a large
variety of somewhat unexpected and exotic final states in the case of fast
quenching. Complementary to this richness it is also interesting to
investigate the (usually less considered) adiabatic
behavior and analyze the factors that govern the corresponding
dynamics.

We find that in general, the dynamics after turning off the interaction
follows basic expectations, i.e. the system reaches a metallic phase.
Notwithstanding, some subtleties arise:
An important (and to our knowledge not yet recognized) issue 
concerns the chemical potential for a generic density of states for which 
particle-hole symmetry does not hold.
In fact, standard protocols for time-dependent interactions in the
BCS model work with a {\it time independent}  chemical potential
which can not adjust to the equilibrium chemical potential of an eventual
final Fermi sea state.
As a consequence, an additional time dependent phase $\phi(t)$ of the order
parameter appears, 
i.e. $\Delta(t)=|\Delta|(t)e^{i\phi(t)}$, that evolves in time while
$|\Delta|\to 0$. Since $\phi$  is the conjugate variable to the charge
we explicitly show that we can use a time-dependent gauge transformation
to eliminate $\phi(t)$ in favor of a new chemical potential $\mu(t)$ which
converges to the  equilibrium chemical potential of the metal. 

Another issue concerns the inverse protocol were one starts from the
metallic state and adiabatically switches on the pairing interaction. 
While in general, as mentioned above, this could be problematic because of
the adiabatic theorem, we will show that the metallic state reached by the
switching-off
procedure is special in that it keeps memory of its superconducting origin.
This ``hidden memory'' is enough to make the dynamics reversible, so that upon
turning on the interaction the system goes back to the superconducting state. 
However, to have a symmetric dynamics requires the application of the superconducting analogous of a an NMR ``$\pi$-pulse'' (for Anderson pseudospins)
at the reversal time. This whole process resembles an NMR Hahn spin-echo/photon echo experiment.~\cite{hahn}

A further issue which we address in the present paper is related
to the gauge invariance of the original BCS theory.
In fact, the BCS wave-function with a fixed phase $\phi$
(hereafter simply  ``BCS'') is deemed to be not-gauge invariant
but the theory can be rendered gauge
invariant upon considering collective modes of the phase.~\cite{and1,and2,gaugin} We analyze this issue in the present setting by studying the role of
$\phi$ in restoring the gauge invariant current response when the
non-interacting Fermi liquid is approached starting from a BCS superconductor
in the presence of an external vector field.

In Sec.~\ref{sec:form} we outline  the formalism starting from a variational approach for the
time-dependent BCS wave-function and discuss some basic questions, e.g.
concerning energy conservation during the adiabatic evolution. 
In the results section~\ref{sec:res}
  we first address the issue of adiabaticity in \ref{sec:ad}, i.e. we answer
  the question how long adiabaticity is preserved upon exponentially
  switching off the pairing interaction and we discuss the question of reversibility. 
 Sec.~\ref{sec:op} then clarifies
  the role of the momentum independent order parameter phase in maintaing
  particle number conservation. The finite momentum phase response is
  then addressed in Sec.~\ref{sec:ed} where we study the time dependent
  response to an applied vector potential both in the transverse and
  longitudinal limit.

\section{Formalism}\label{sec:form}
The superconducting system is described within the attractive Hubbard model
\begin{equation}
  H=\sum_{ij,\sigma}(t_{ij}-\mu_0\delta_{ij}) c_{i,\sigma}^\dagger c_{j,\sigma}
  -U(t)\sum_i n_{i,\uparrow}n_{i,\downarrow}\label{eq:hi}
\end{equation}
where the first term corresponds to the kinetic energy of
electrons on a lattice (hopping amplitude $t_{ij}$ between lattice
site $R_i$ and $R_j$) and 
$-U(t)<0$ denotes a time-dependent local attraction.
Specifically, we consider an exponential switching off
\begin{equation}\label{eq:ut}
  U(t)=U e^{-t/T}
\end{equation}
which is parametrized by the time-scale $T$.

\subsection{Variational Dynamics}  

The dynamics is evaluated variationally by means of a
time-dependent BCS wave function \cite{bcs}
\begin{equation}\label{eq:varbcs}
  |\Psi_{BCS}(t)\rangle = \prod_{k}\left\lbrack u_k(t) + v_k(t) e^{i\phi(t)}c_{k,\uparrow}^\dagger c_{-k,\downarrow}^\dagger \right\rbrack |0\rangle
\end{equation}
where we have specified the relative phase $\phi$ of $u_k$ and $v_k$ such
that $\sum_k u_k v_k^*$ is real. In fact, if $\sum_k u_k' (v_k')^*$ were
not real, we can always make the change of variables $v_k=v_k'e^{-i\phi(t)}$
with $\phi(t)$ chosen such that the new sum of products is real.

The variational solution of the time-dependent Schr\"odinger  equation can be
obtained by requiring the action $S=\int dt L $ to be stationary with the following real Lagrangian \cite{Blaizot1986}
\begin{equation}
L = \frac{i}{2} \frac{\langle \Psi_{BCS}|\dot{\Psi}_{BCS}\rangle 
- \langle \dot{\Psi}_{BCS}|{\Psi}_{BCS}\rangle}{\langle\Psi_{BCS}|\Psi_{BCS}\rangle}
- \frac{\langle\Psi_{BCS}| H |\Psi_{BCS}\rangle}{\langle\Psi_{BCS}|\Psi_{BCS}\rangle}
\end{equation}
which leads to the equations of motion from the standard Euler-Lagrange 
equations.

Evaluation of the Lagrangian yields
\begin{eqnarray}
  L&=&\frac{i}{2}\sum_k\left(u_k^*\dot{u}_k-u_k\dot{u}_k^*+v_k^*\dot{v}_k-v_k\dot{v}_k^* \right)\nonumber \\ &-&N \dot{\phi}\frac{n}{2}- E_{BCS} \label{eq:lag}
\end{eqnarray}
with
\begin{equation}
  E_{BCS}=2\sum_{k}(\varepsilon_k-\mu_0)|v_k|^2
  - N U(t) \left\lbrack \frac{n^2}{4}+ |f|^2  \right \rbrack \label{eq:ebdg}
\end{equation}
where $\varepsilon_k=1/N\sum_{nm} t_{nm} e^{ik(R_n-R_m)}$ denotes the electronic
dispersion and $N$ is the total number of lattice sites.
We have also defined density $n$ and Gorkov function $f$ as
\begin{eqnarray}
n &=& \frac{2}{N}\sum_k |v_k|^2 \\
f &=& e^{i\phi} \frac{1}{N}\sum_k u_k^* v_k  \, .\label{eq:gork}
\end{eqnarray}
It follows from our definition of $|\Psi_{BCS}(t)\rangle$ that the phase of
$f$ is completely specified by $\phi$.

The dynamics can be conveniently solved by defining the density matrix
\begin{eqnarray}
\underline{\underline{R}}(k) &=& \left(
\begin{array}{cc}
R_{11}(k) & R_{12}(k) \\
R_{21}(k) & R_{22}(k)
\end{array}\right)
= \left(
\begin{array}{cc}
|v_k|^2 & u_k v_k^* e^{-i\phi} \\
u_k^* v_k e^{i\phi} & 1-|v_k|^2
\end{array}\right) \nonumber\\
&=&
\left(
\begin{array}{cc}
\langle c_{k,\uparrow}^\dagger c_{k,\uparrow}\rangle_{BCS}  & \langle c_{k,\uparrow}^\dagger c_{-k,\downarrow}^\dagger\rangle_{BCS} \\
\langle c_{-k,\downarrow}c_{k,\uparrow}\rangle_{BCS} &
\langle c_{-k,\downarrow}c_{-k,\downarrow}^\dagger\rangle_{BCS}
\end{array}\right) \label{eq:dm}
\end{eqnarray}
which obeys the equation of motion
\begin{equation}
  \frac{d}{dt}\underline{\underline{R}}(k)=-i\left\lbrack
  \underline{\underline{R}}(k),H^{BCS}(k)\right\rbrack \label{eq:densmat}
  \end{equation}
and the BCS hamiltonian is evaluated from
$H^{BCS}_{nm}(k)=\partial E_{BCS}/\partial R_{mn}(k)$.

It is also convenient to introduce the spinors
\begin{eqnarray}
 J_{k}^x &=&  \frac{1}{2}\langle c_{k,\uparrow}^\dagger c_{-k,\downarrow}^\dagger+c_{-k,\downarrow} c_{k,\uparrow}\rangle_{BCS}, \label{eqjx}\\
J_{k}^y &=& -\frac{i}{2}\langle c_{k,\uparrow}^\dagger c_{-k,\downarrow}^\dagger-c_{-k,\downarrow} c_{k,\uparrow}\rangle_{BCS} , \label{eqjy}\\
J_{k}^z &=& \frac{1}{2}\langle 1- c_{k,\uparrow}^\dagger c_{k,\uparrow}-c_{-k,\downarrow}^\dagger c_{-k,\downarrow}\rangle_{BCS}  \label{eqjz}
\end{eqnarray}
so that the equations of motion can be written as
\begin{equation}\label{eq:jmag}
  \frac{d}{dt}{\bf J}_k = 2 {\bf b}_k \times {\bf J}_k
\end{equation}
with an effective 'magnetic' field
\begin{equation}\label{eq:mag}
  {\bf b}_k = \left(
  \begin{array}{c}
    -U f_x \\
    -U f_y \\
    \xi_k
  \end{array}\right)\,.
\end{equation}
Here we have defined $\xi_k\equiv \varepsilon_k-\mu_0-\frac{U n}{2}$
and the Gorkov function $f=f_x-i f_y=1/N \sum_k (J_k^x-i J_k^y)$.
The BCS order parameter is then given by $\Delta=-U f$.

The time dependence of the total energy $E_{BCS}(t)$ follows from
\begin{eqnarray}
  \frac{d E_{BCS}}{dt}&=&\sum_{k}\frac{\partial E_{BCS}}{\partial R_{nm}(k)}\dot{R}_{nm}(k) -N\frac{\partial U(t)}{\partial t}\left\lbrack |f|^2 +\frac{n^2}{4}\right\rbrack\nonumber\\
  &=& -i \sum_{k} Tr\left\lbrace \underline{\underline{H}}^{BCS}(k)
  \left\lbrack \underline{\underline{R}}(k), \underline{\underline{H}}^{BCS}(k)\right\rbrack \right\rbrace \nonumber \\
  &-& N \frac{\partial U(t)}{\partial t} \left\lbrack |f|^2+\frac{n^2}{4}\right\rbrack \nonumber\\
  &=& -N \frac{\partial U(t)}{\partial t} \left\lbrack |f|^2 +\frac{n^2}{4}\right\rbrack \label{eq:econs}
\end{eqnarray}
where the term in the second line vanishes upon permutating the
trace. For an interaction which is constant in time the energy is
therefore conserved.

Equation~(\ref{eq:econs}) is the BCS approximated version
of the exact equation
\begin{equation}\label{eq:denerdt}
  \frac{dE}{dt}=-N \frac{\partial U(t)}{\partial t}\langle n_\uparrow n_\downarrow  \rangle
\end{equation}
which follows from the Heisenberg equation of motion applied to
Eq.~(\ref{eq:hi}). For an adiabatic turning-off of the interaction one has
${dE}/{dt}>0$. This means that the total energy increases on going from the superconducting state to the normal state as expected. 
In this context $U(t)$ can be considered as an external ``drive'' so that the
increase in energy corresponds to the work exerted by the drive on the system.

The equations of motions Eq.~(\ref{eq:jmag}) can also be derived
by taking as a starting model a BCS Hamiltonian
\begin{equation}\label{eq:hbcs}
  H=\sum_k \xi_k J_k^z-\frac{U(t)}{N_{norm}}\sum_{p,k}J_p^+J_k^-
\end{equation}
where $k$ and $p$ are energy labels,  $J_p^+=J_p^x\pm iJ_p^y$
and  $N_{norm}$ is a normalization
factor given by the number of levels.  With a constant $U(t)=U_0$
the model becomes integrable in the thermodynamic limit, leading then to a constraint dynamics,
which reflects in the peculiar properties of the model
under quenching~\cite{bara04,yus05,bara06,alt06,yus06,Scaramazza2019} or driving.~\cite{Collado2018} 
These considerations upgrade Eqs.~(\ref{eq:ebdg})-(\ref{eq:econs}) from approximate to exact equations for the dynamics of model Hamiltonian Eq.~(\ref{eq:hbcs}) in the thermodynamic limit. 


\subsection{The equilibrium chemical potential problem}  
We know discuss the issue related to the chemical potential mentioned in the
introduction. 
The chemical potential, which appears in Eq.~(\ref{eq:hi}) and 
enters the definition of $\xi_k$
is the equilibrium value $\mu_0$ related to the initial state with
$U_0=U(t=0)$. This quantity influences the dynamics via the $z$-component
of the vector ${\bf b}_k$ in  Eq.~(\ref{eq:mag}).

$\mu_0$ is not a dynamical variable but a parameter in the Hamiltonian so  it stays constant in the equation of
motion Eq.~(\ref{eq:jmag}). On the other hand,  in general, after a generic change of parameters the chemical potential that yields the same number of particles in the grand canonical sense, i.e. after minimization with respect to $N$,
will be different. Furthermore,  the total number of
particles is conserved by Eqs. (\ref{eq:jmag}, \ref{eq:mag}) since
$ d/dt \sum_k J_k^z=0$, (cf. below) so keeping the ``wrong'' chemical potential does not change the number of particles.  Still it would be useful to find a procedure which at the end of the dynamics yields the
``correct'' grand canonical chemical potential compatible with the fixed number of particles. Actually we will see that one can define a 
grand canonical $\mu(t)$ at all times during the adiabatic evolution.
We shall see that the solution to this problem  relies on a gauge transformation. Notwithstanding that, the gauge invariant physical quantities can be
derived from Eq.~(\ref{eq:jmag}) with the constant chemical potential, $\mu_0$.

\section{Results}\label{sec:res}
\subsection{Adiabatic dynamics}\label{sec:ad}
All calculations are done for
a semielliptic density of states (DOS)
$\rho(\omega)=\frac{2}{\pi}\sqrt{B^2-\omega^2}$ with $B\equiv 1$ which we
discretize into $10.000$ intervals (but for some results in Sec.~\ref{sec:inv}).
We have checked that results do not change upon further increasing the
number of energy intervals.

\begin{figure}[h!]
\includegraphics[width=7.5cm,clip=true]{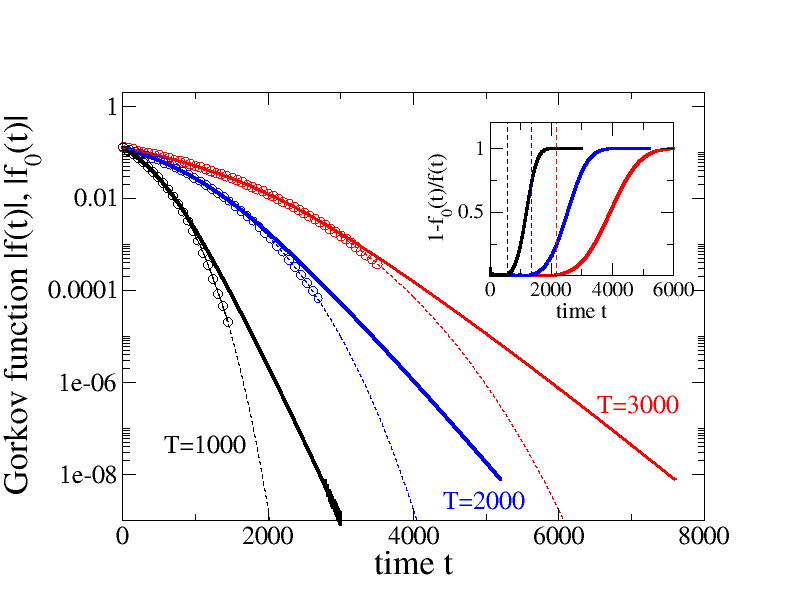}
\caption{ Time evolution of the Gorkov function $|f(t)|$ (solid lines) for
  different adiabatic time scales $T=1000$ (black), $T=2000$ (blue), and $T=3000$ (red). Circles correspond to the equilibrium values $|f_0(t)|$
  obtained from the BCS equation for the interaction value $U(t)$
  from Eq.~(\ref{eq:ut}) and the dashed lines are the corresponding parametrization from Eq.~(\ref{eq:fitbcs}). The inset shows $1-f_0(t)/f(t)$ for the different adiabatic time scales in comparison with the time where the adiabaticity
  condition $d\Delta/dt<\Delta^2$ starts to get violated (dashed).
  Results are for a semielliptic DOS $\rho(\omega)=\frac{2}{\pi}\sqrt{B^2-\omega^2}$ with $B\equiv 1$ at density $n=0.875$ and an initial value of $U(t=0)=0.5$. } 
\label{fig:1}                                                   
\end{figure}

Figure~\ref{fig:1} shows the dynamics of the (magnitude of the) Gorkov
function $|f(t)|$ for different adiabatic time scales $T$, cf.
Eq.~(\ref{eq:ut}) and $U(t=0)=0.5$. 
For comparison we also show the equilibrium values $f_0$ obtained from
the BCS equation for the corresponding interaction values $U(t)$, cf.
Eq.~(\ref{eq:ut}), which for $U(t=0)=0.5$ can be fitted by the
BCS equation
\begin{equation}\label{eq:fitbcs}
  |f_0(t)|=\frac{\alpha}{U(t)}e^{-\frac{1}{N_0 U(t)}}
\end{equation}
with effective energy scale $\alpha=1.45$ and effective DOS $N_0=1.575$.
With increasing adiabatic decay time $T$ the
time dependent Gorkov function $f(t)$ increasingly coincides with $f_0$.
The regime of adiabaticity scales with $T$ while beyond this
regime $f(t)$ overestimates the equilibrium value
due to incomplete relaxation, cf. below. The inset to Fig.~\ref{fig:1}
illustrates the violation of adiabaticity for $T<T_{ad}$ as indicated
by the deviation of $1-f_0/f$ from zero.
The adiabatic time scale can also be defined via the equation \cite{polkov11},
\begin{equation}
  \frac{d\Delta}{dt}|_{t=Tad}=\Delta^2(T_{ad})
\end{equation}
from which we have numerically extracted $T_{ad}$ 
by comparing the rate of change of the  gap with the square of the gap
{\it vs.} time.
The corresponding values of $T_{ad}$ (dashed lines
in the inset to Fig.~\ref{fig:1}) show a remarkable agreement with
$T_{ad}$ as determined from the time where  $f_0(t)$ and $f(t)$ start diverging from each other.


\begin{figure}[hhh]
\includegraphics[width=8.5cm,clip=true]{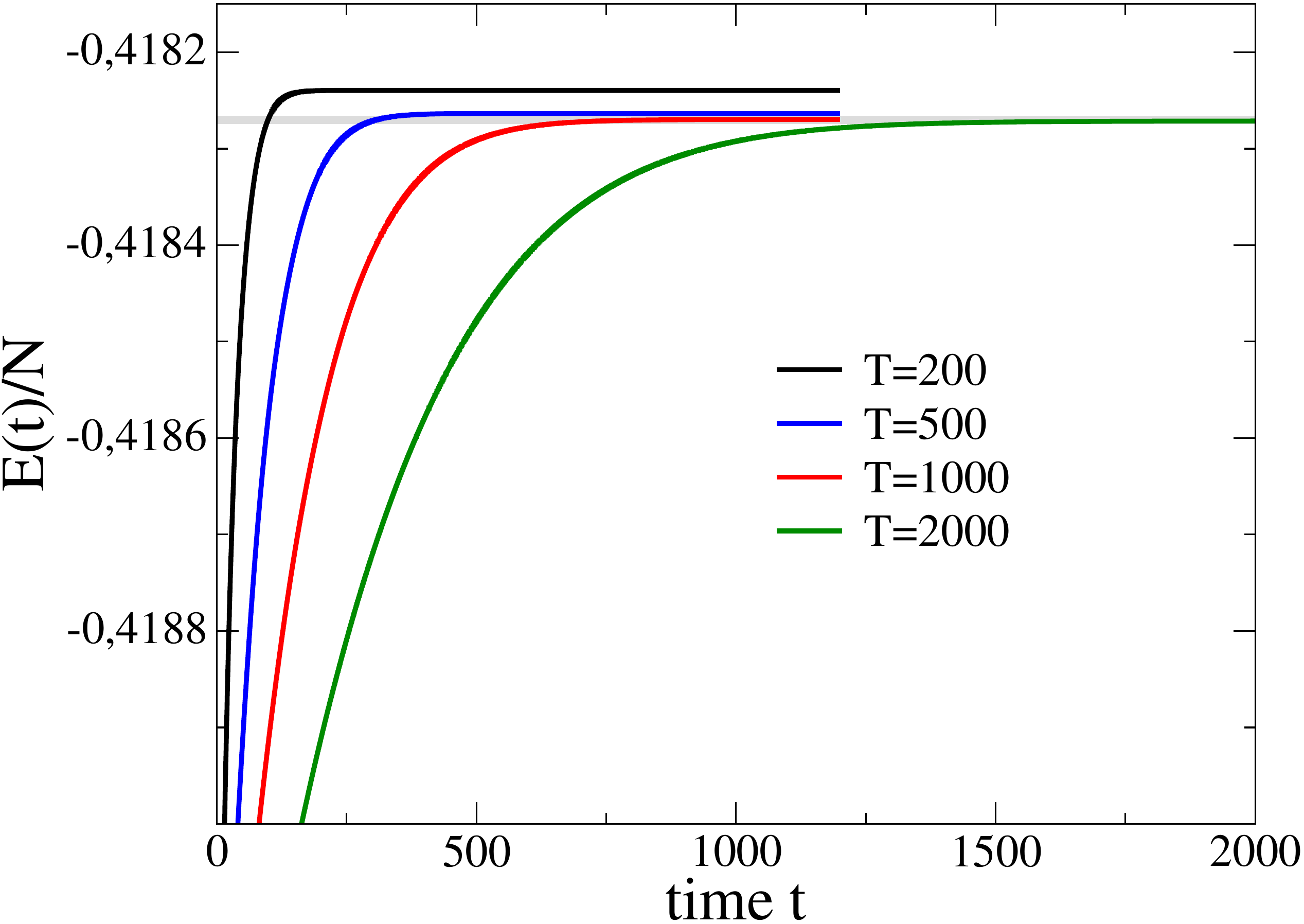}
\caption{Time dependence of the energy per site for different
  adiabatic time scales.
  The horizontal grey is the equilibrium value for $U=0$.
System and parameters
are the same than in Fig.~\ref{fig:1}.
} 
\label{fig:2}                                                   
\end{figure}

Figure~\ref{fig:2} shows the time evolution of the total
energy. For large enough adiabatic decay times $T$
the final energy approaches the
equilibrium value for $U=0$, whereas this is overshot for small $T$.
One can shed some light on this behavior by noticing that $U(t)$, as defined in Eq. (\ref{eq:ut}), is invertible so that the time derivative
in Eq.~(\ref{eq:econs}) can be transformed into a derivative in $U$, i.e.
\begin{equation}
  \frac{dE}{dU}=-N\left( \frac{n^2}4+|f|^2 \right)\,.
\end{equation}
Then, it follows that the energy difference $\Delta E(t) $ between interacting
system at time $t=0$ and the final system at time $t$ is given by
\begin{equation}\label{eq:cmp}
  \Delta E(t) =-N\int_{U_0}^{U(t)} dU \left( \frac{n^2}4+|f|^2 \right) = \int_0^t \frac{dE}{dt}dt
\end{equation}
where in a complete adiabatic process the Gorkov function can be replaced
by its equilibrium value $f_0$.

For small decay times $T$, the dynamic Gorkov function $f(t)$ overestimates
the adiabatic equilibrium value already at an early stage of the
time evolution leading therefore also to an overestimation of $\Delta E(t\to\infty)$. This difference between $f(t)$ and $f_0$ decreases with
increasing $T$ and for $T\gtrsim 1000$ the difference between $E(t)$ and
the energy of the non-interacting Fermi liquid cannot be resolved on the
scale of the plot.

\begin{figure}[hhh]
\includegraphics[width=8.5cm,clip=true]{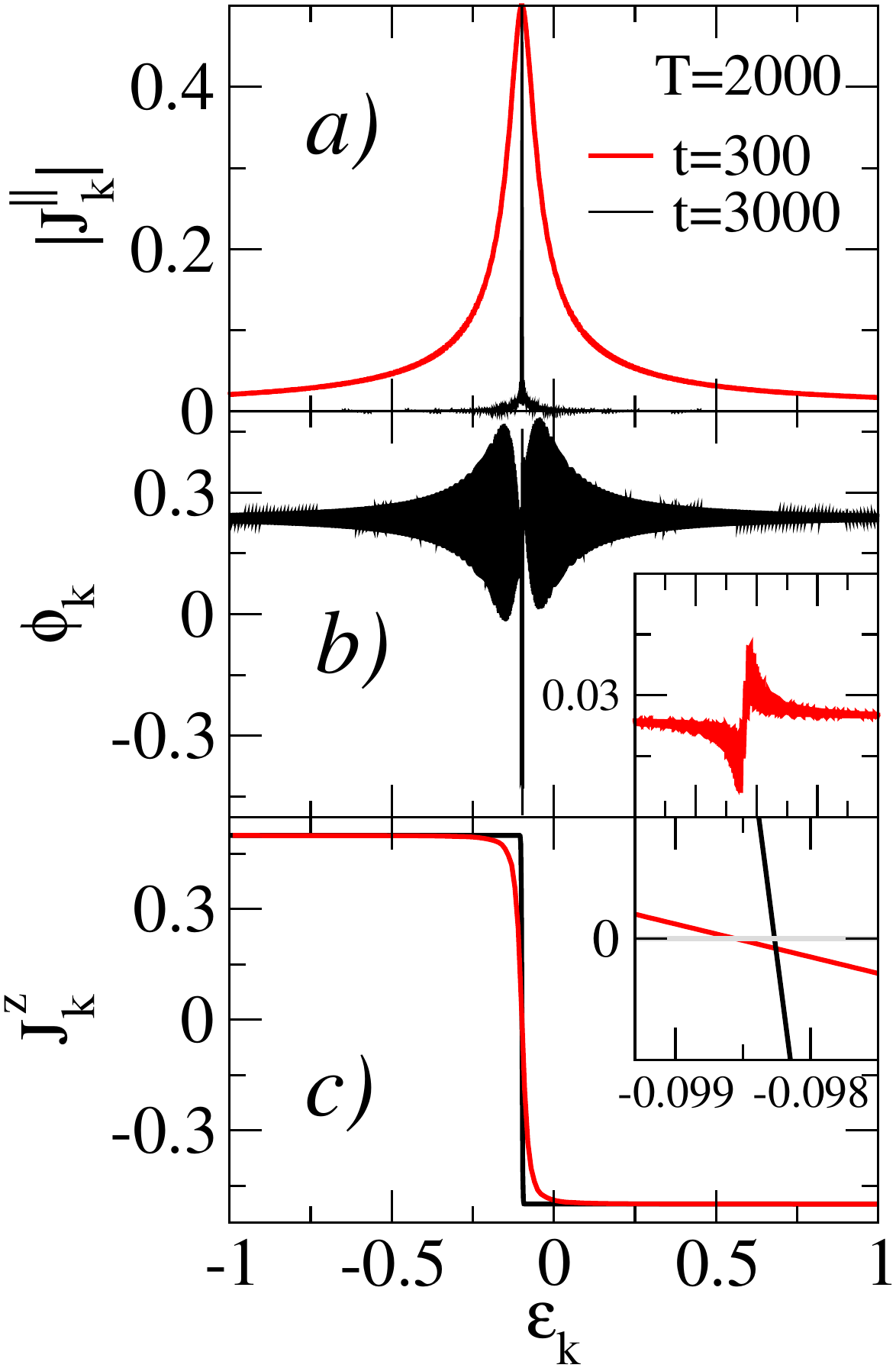}
\caption{Dependence of $J_k^\parallel\equiv\sqrt{(J_k^{x})+(J_k^{y})}$ (a),
  the momentum dependent phase $\phi_k=\arctan( J_k^{y}/J_k^{x})$ [main panel (b) and inset] 
 and $J_k^z$ (c) 
  on $\varepsilon_k$ at times
  $t=300$ and $t=3000$ for an adiabatic decay time $T=2000$.
  The inset to (c) details the behavior of $J_k^z\approx 0$.
Results are for a semielliptic DOS $\rho(\omega)=\frac{2}{\pi}\sqrt{B^2-\omega^2}$ with $B\equiv 1$ at density $n=0.875$ and an initial
value of $U(t=0)=0.5$.
} 
\label{fig:3}                                                   
\end{figure}

The energy dependence of the pseudospins $J_k^{x,y,z}$, Eqs. (\ref{eqjx}, \ref{eqjy}, \ref{eqjz}), for two
times $t=300$ and $t=3000$ and interaction decay time $T=2000$ is shown in Fig.~\ref{fig:3}.
In the BCS approximation these quantities are given by
\begin{eqnarray}
  J_k^x&=&-\frac{1}{2}\frac{|\Delta|\cos(\phi)}{\sqrt{|\Delta|^2+\xi_k^2}}\label{eq:jkx} \\
  J_k^y&=&\frac{1}{2}\frac{|\Delta|\sin{\phi}}{\sqrt{|\Delta|^2+\xi_k^2}} \label{eq:jky}\\
  J_k^z &=& -\frac{1}{2}\frac{\xi_k}{\sqrt{\Delta^2+\xi_k^2}}\label{eq:jkz}
\end{eqnarray}
where $\phi$ was defined in Eq.~(\ref{eq:gork}).

Since we are in the adiabatic regime, the pseudospins at $t=300$ are
fully compatible with the BCS result when evaluated with the gap $\Delta(t=300)$
and phase $\phi(t=300)$ at the same time. However, minute deviations
of the individual phases $\phi_k=\arctan( J_k^{y}/J_k^{x})$ from the
global phase
$\Phi\approx 0.03$ are already apparent as shown in the
inset to Fig.~\ref{fig:3}.
The issue of a finite global phase will be
discussed in Sec.~\ref{sec:op}.

\begin{figure}[hhh]
\includegraphics[width=8.5cm,clip=true]{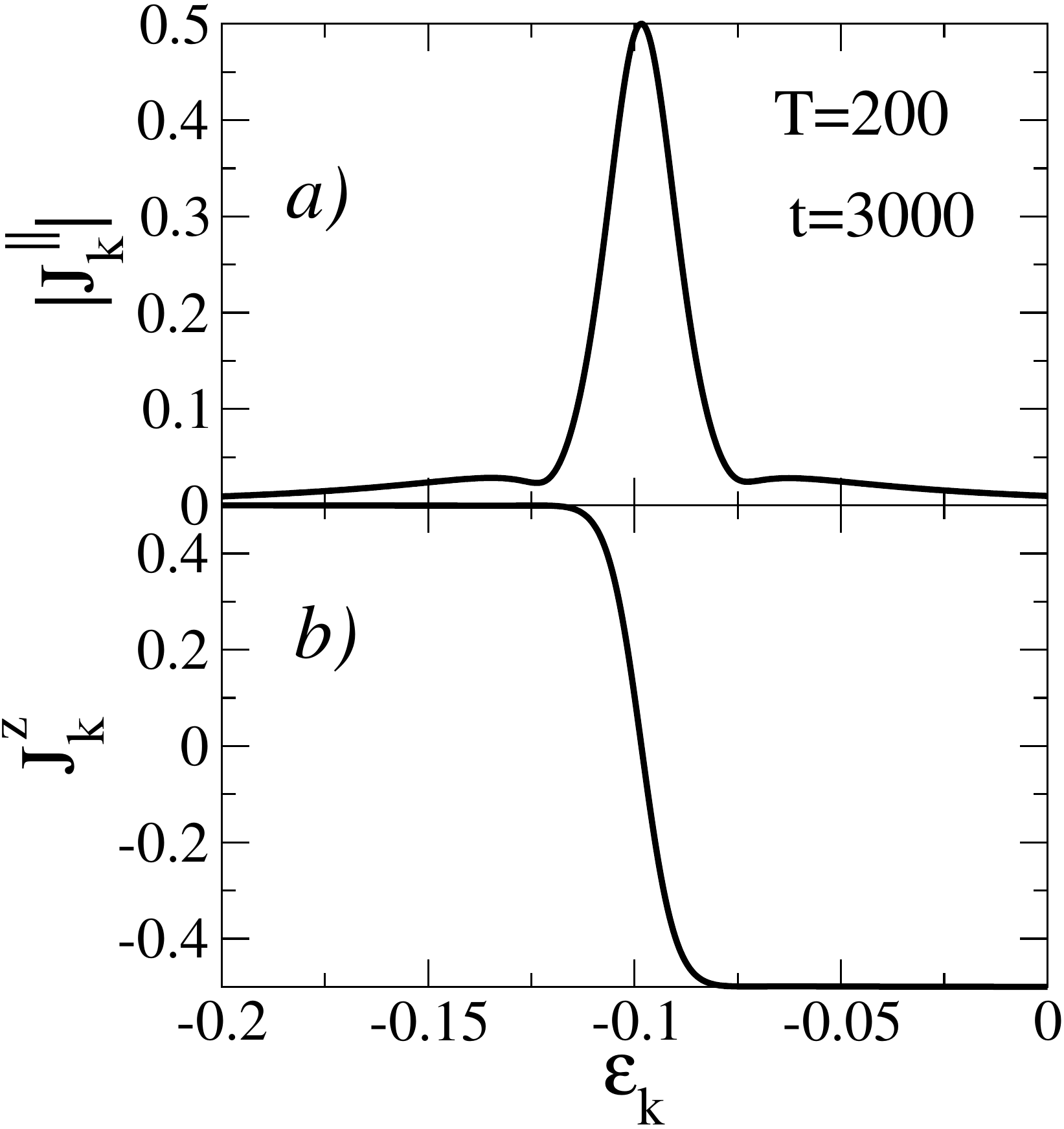}
\caption{Same as panels (a,c) of Fig. \ref{fig:3} but for a
  small adiabatic decay time $T=200$ and $t=3000$.} 
\label{fig:app1}                                                   
\end{figure}

At large times $t=3000$ the gap tends to zero and $J_k^z$, Eq.~(\ref{eq:jkz})
develops into a step function which in the present context of adiabatic
behavior is nothing but the Fermi function at zero temperature.
As can be seen from the inset to Fig.~\ref{fig:3}c, the energy at
which $J_k^z$ equals zero changes with the time evolution of the system.
In Sec.~\ref{sec:op} we will discuss how this shift is related to the change
in the chemical potential. Far from the adiabatic limit, i.e. for $t\gg T$ (cf. Fig.~\ref{fig:app1}), $J_k^z$ does not completely evolve into a step function
but resembles a finite temperature Fermi function, though the functional
form is different.
This is of course related to the incomplete relaxation and the remaining
excitations are responsible for the extended finite slope in $J_k^z$ around
$\varepsilon_k=\mu$.

Since the length of the pseudospins $|J_k^x|^2+|J_k^y|^2+|J_k^z|^2=1/4$ is
conserved, a zero in $J_k^z$ implies maximum projection of $\vec{J}_k$ onto the other two components. Indeed, one finds that as a function of energy, the corresponding magnitude
$J_k^\parallel=\sqrt{(J_k^x)^2+(J_k^y)^2}$ develops into a narrow spike at
large times, cf. Fig. \ref{fig:3}a. 
For $T=300$ this behavior can be understood as due to the residual gap,
however, even for infinite times a hypothetical pseudospin at the position
of the discontinuity in  $J_k^{z}$ (i.e.
at $J_k^{z}=0$) should remain out of equilibrium in such a way
that $J_k^\parallel=1/2$.

Of course this depends on whether the state $\xi_k=0$ is an accessible state
of the system or not. Thus, we first consider a {\it finite} system
for which $\xi_k=0$ is {\it not} an accessible state in the sense that during
the whole time evolution the chemical potential remains localized between two 
levels.

Then the density matrix for $t\to\infty$ takes the form
\begin{eqnarray}
\underline{\underline{R}}_\infty&=& \frac12 (\underline{\underline{1}}
+\underline{\underline{\tau_z}}) \hspace*{0.3cm}\mbox{for}\hspace*{0.3cm} \xi_k<0 \nonumber \\
\underline{\underline{R}}_\infty&=& \frac12 (\underline{\underline{1}}
-\underline{\underline{\tau_z}})\hspace*{0.3cm} \mbox{for}\hspace*{0.3cm} \xi_k>0 \label{eq:dens0}
\end{eqnarray}
so that upon starting from a BCS superconductor and adiabatically switching
off the interaction the system approaches a normal Fermi liquid
with all states below (above)
$\xi_k=0$ occupied (empty).

In the opposite limit, i.e. by first taking the large $N$ limit and
then evaluating the dynamics for large times, the system always
keeps memory of its initial superconducting
state when approaching the Fermi liquid as there are always states
sufficiently close to the sign change of $J_k^z$
which have a non-zero pseudospin component in the $xy$-plane.
So in case of a decreasing interaction, no matter how slow and
how small the final interaction is, 
the dynamics can be always inverted (see next subsection)
at a given time and one
returns to the superconducting state.

In the other case, i.e. if $\xi_k=0$ is an accessible state, then according to the above discussion, the density matrix for this level and in the limit $t\to\infty$ reads
\begin{equation}
\underline{\underline{R}}_\infty=\frac12 (\underline{\underline{1}}
+\cos(\varphi)\underline{\underline{\tau_x}}-\sin(\varphi)\underline{\underline{\tau_y}})\hspace*{0.3cm} \mbox{for} \hspace*{0.3cm}\xi_k=0 \,.
 \label{eq:dens1}
\end{equation}

It can be shown \cite{Blaizot1986} that the density matrix describes a pure
BCS state if and only if it is idempotent, i.e. $ \underline{\underline{R}}^2=\underline{\underline{R}}$. One can check that this is the case
for the above density matrix. Thus, the pseudospin at $\xi_k=0$ keeps
track of the BCS origin of the state and $\varphi$ keeps memory of the original
phase $\phi$ of the state. In Appendix \ref{a2} it is shown that
such an initial state with only one transverse pseudospin component
at $\xi_k=0$ evolves into a  BCS state upon turning on the attractice
interaction. However, the BCS recovery time scales with the system size
so that in the thermodynamic limit the system stays inifinitely long in this
unstable saddle point.

Notice that for a metal the density matrix at the chemical potential will
normally be taken as $\underline{\underline{R}}=\frac12 \underline{\underline{1}}$ which is manifestly non idempotent. Indeed this density matrix does
not correspond to a BCS state nor to a single Slater determinant
(which requires $n_k^2=n_k$) but needs a mixed state to be represented.

\begin{figure}[hhh]
\includegraphics[width=8.5cm,clip=true]{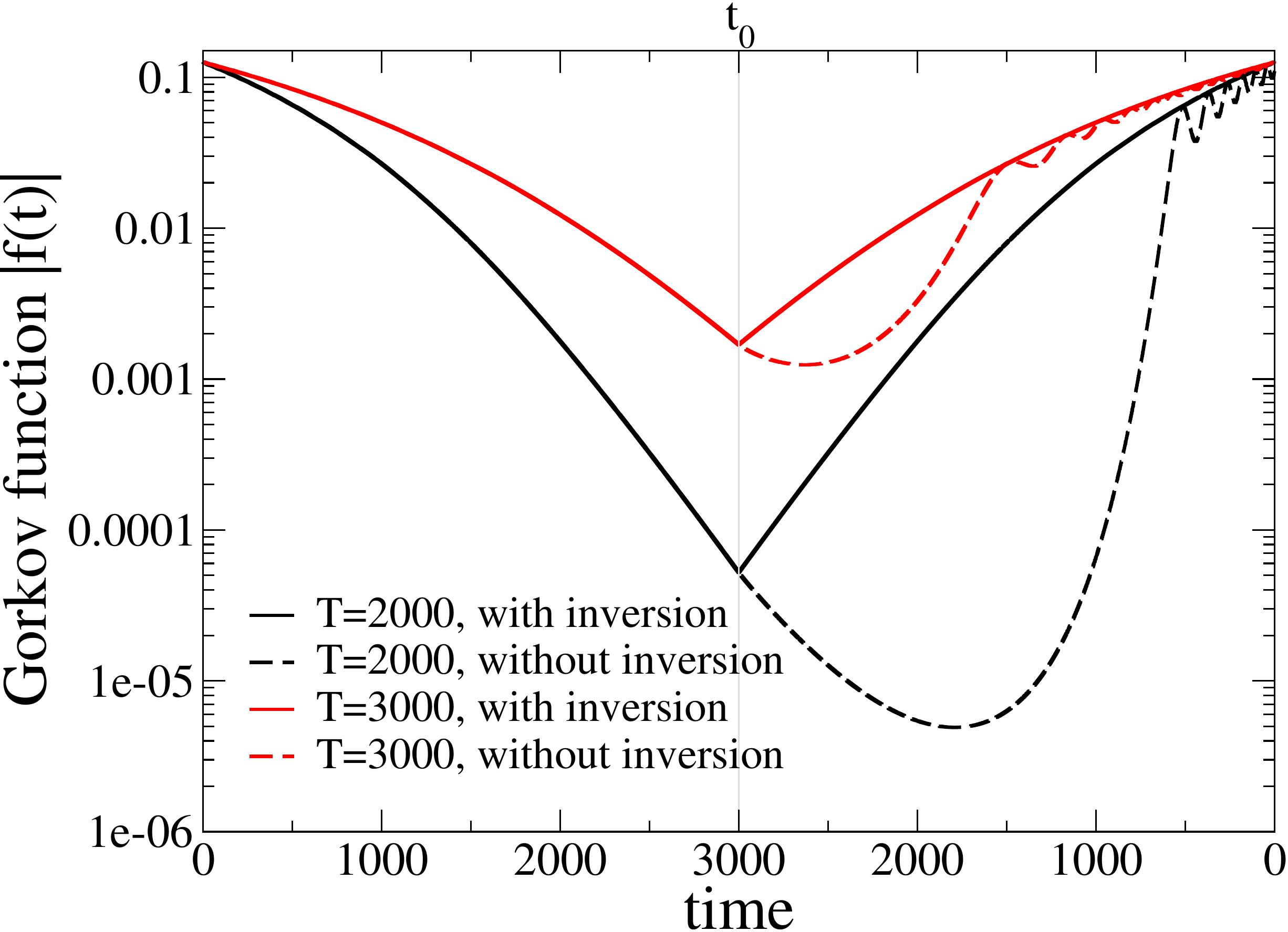}
\caption{Time dependence of the Gorkov function where the
  time evolution is reversed after $t_0=3000$ time steps with adiabatic
  time scales $T=2000$ (black) and $T=3000$ (red). Solid (dashed)
  lines correspond to the reversed time evolution with (without) inversion
  of the $J_k^{x,y}$ component. 
  Interaction parameter $U(t=0)=0.5$; density $n=0.875$.} 
\label{fig:3a}                                                   
\end{figure}  

\subsection{Time inversion of the dynamics }\label{sec:inv}

We now discuss the issue of inverting the dynamics after some given time $t_0$
where we will show that this requires the application of
the equivalent of a $\pi$-pulse in an NMR Hahn echo experiment.~\cite{hahn}
In fact,  inspection of the equations of motion Eq.~(\ref{eq:jmag})
reveals that $t\to -t$ implies $J_k^x \to -J_{k}^x$ {\it or} $J_k^y \to -J_{k}^y$ (and thus $f_x \to -f_x$ {\it or} $f_y \to -f_y$).
The time inversion protocol is therefore specified as follows:
\begin{enumerate}
\item Evaluate the dynamics of the system with a decreasing interaction $U(t)=U e^{-t/T}$
  until a given time $t=t_0$.
\item At $t=t_0$ perform the rotation $J_k^x \to -J_{k}^x$ {\it or} $J_k^y \to -J_{k}^y$.
\item At $t=t_0$ transform to an increasing
  interaction $U(t)=U(t_0)e^{(t-t_0)/T}$ so that the total $U(t)$ is symmetric
  with respect to $t_0$ and the initial interaction is recovered
  at $t=2t_0$, i.e. $U(t=0)=U(2t_0)=U$.
\item Evaluate the dynamics of the system which at $t=2t_0$ approaches the
  initial state.
\end{enumerate}

Fig.~\ref{fig:3a} reveals that inversion of time  with the concomitant
inversion of one transverse pseudospin component, indeed, leads to symmetric
behavior of the Gorkov function with respect to the inversion point at $t=t_0$.
On the contrary, without inversion of $J_k^{x,y}$ the Gorkov function
keeps decreasing after the time inversion thus showing an intrinsic inertia
in some restricted time interval. Such inertial behavior can be understood
by mapping the pseudospin dynamics to harmonic oscillators.\cite{OjedaCollado2021} In the present setting, for a negligible $\Delta$, the $J_k^{x}$ and $J_k^{y}$ components play the role of the canonical variables describing circles in phase space. The flip of   $J_k^{y}$ corresponds to a sign change of the velocity at the inversion point. Without such inversion, the inertia tends to keep
the original trajectory until the external drive takes over.

Notice that the above protocol keeps both phase and amplitude information,
justifying our previous statement that the individual
pseudospin phase at the
chemical potential keeps information of the global phase of the parent
superconductor. 

Literally, the $\pi$-pulse NMR protocol to  
send $J_k^y \to -J_{k}^y$ would require a $\pi$-rotation along the
$x$ direction which can be achieved here by a time-dependent $b_k^x$ due to a
controllable Josephson coupling with another superconductor (or band).
In addition, the field should be applied  only to a small set of
pseudospins close to the chemical potential in order to minimize the
simultaneous introduction of charge fluctuations ($J_k^z \to -J_{k}^z$).
In the present computation we introduced the $J_k^y $ flip ``by hand''
at the inversion time $t_0$, leaving for future work possible practical implementations by generalized external fields.

\subsection{Global phase dynamics and chemical potential }\label{sec:op}
The results of Sec.~\ref{sec:ad}, in particular
energy $E(t)$ (Fig.~\ref{fig:2}) and  $J_k^z(t)$
(Fig.~\ref{fig:3}) are evaluated using Eqs. (\ref{eq:jmag}, \ref{eq:mag})
with $\mu_0 = const$.  Indeed, $E(t)$ and $J_k^z(t)$ are
gauge invariant quantities which do not depend on the phase $\phi$ of the
order parameter.
The phase $\phi$ of the order parameter is the conjugate variable to
the charge density $n$ as can be seen from the Lagrangian Eq.~(\ref{eq:lag}).
Since $E_{BCS}$ does not depend on $\phi$ the corresponding equation
of motion $d/dt(\partial L/\partial\dot{\phi})=-N/2 dn/dt=0$ implies
particle number conservation during the dynamics.

We can make a gauge transformation that eliminates the 
 time-derivative of  $\phi$ in favor of a scalar potential that can be absorbed in the chemical potential
so that in the new gauge  $\phi$ is constant and $\mu$ becomes time-dependent.  
From the term with the time derivative of the phase in  Eq.~(\ref{eq:lag}) it is clear that the 
gauge transformation is $\mu_0\to \mu(t)$ with
\begin{equation}\label{eq:larm} 
  \mu(t)= \mu_0-1/2 d\phi/dt \,.
\end{equation}
This corresponds to a change of phase of the fermionic operator
\begin{equation}\label{eq:gt}
  c_{k,\sigma} \to c'_{k,\sigma} e^{i\phi/2}
\end{equation}
and at the same time the order parameter, transformed in the same way, 
becomes a real quantity.

Since the scalar potential appears as a pseudomagnetic field in the $z$ direction, the same result can be obtained by  transforming the dynamics
to the global Larmor frame of the precessing pseudospins
\begin{displaymath}
  \frac{{d\bf J}_k}{dt} \longrightarrow \frac{{d\bf J'}_k}{dt}+{\bf \Omega}\times {\bf J'}_k \,.
\end{displaymath}
with ${\bf \Omega}=(0,0,d\phi/dt)$ and ${\bf J'}_k $ given by expressions (\ref{eqjx})-(\ref{eqjz}) but in terms of the transformed fermion operators. 
This is in fact equivalent to applying the gauge
transformation Eq.~(\ref{eq:gt}) which eliminates the phase of the
Gorkov function $f$ and which therefore becomes purely real, i.e. $f'_y=0$.

\begin{figure}[h!]
\includegraphics[width=8.5cm,clip=true]{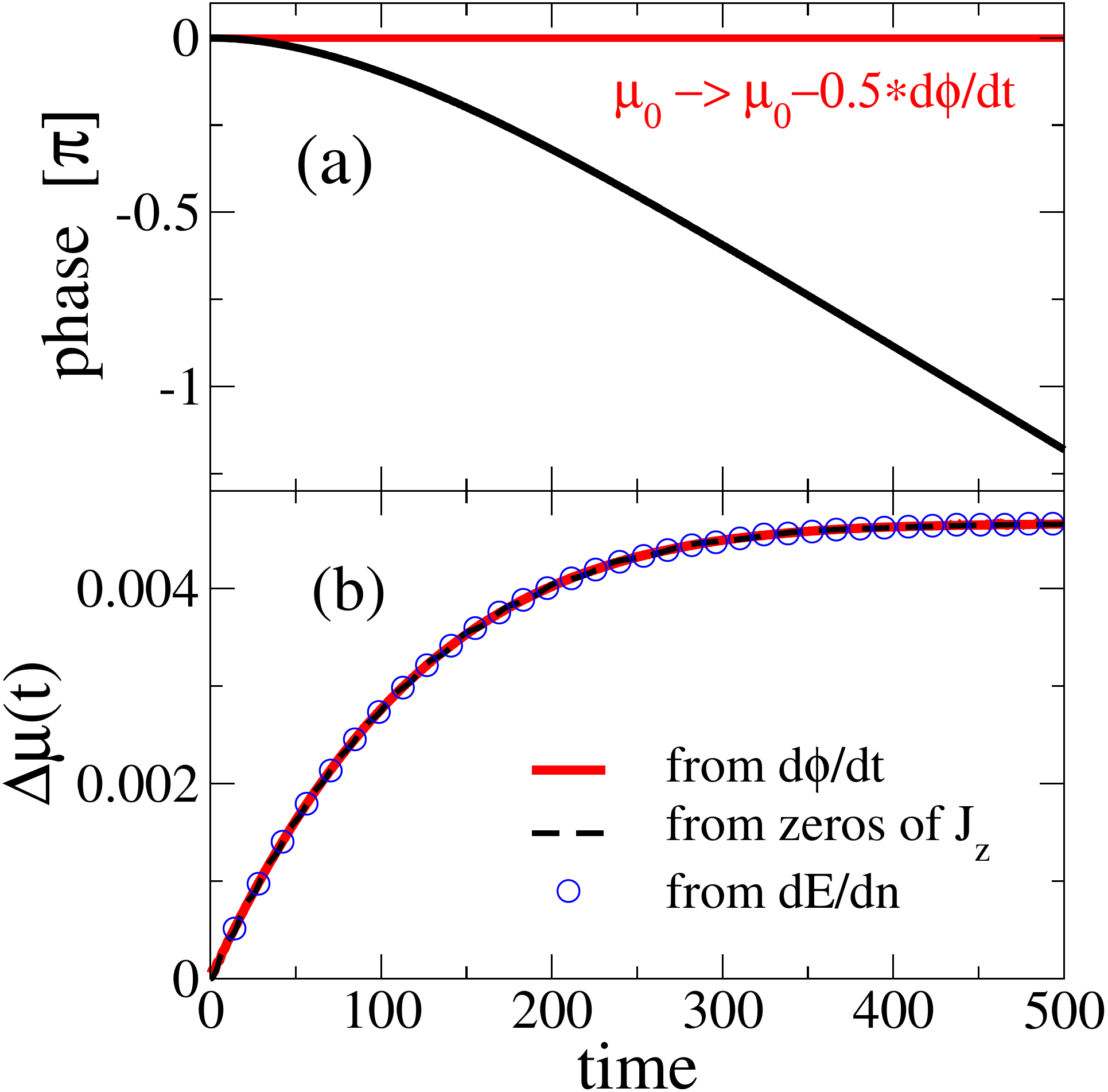}
\caption{(a)  
Phase of the Gorkov function for the
  case $\mu=const.$ (black) and after transforming to the Larmor frame
  (red). (b) Time-dependent chemical potential from transforming to the Larmor frame (red), from the zeros of $J_z$ (black) and from the relation
  $\mu(t)=dE(n,t)/dn$ (blue). Density $n=0.875$, $U/t=1$, decay time $T=500$.}
\label{fig:4}                                                   
\end{figure}  

From a physical perspective this means that in order to follow an adiabatic
dynamics, where at each instant of time one can obtain the solution
from the BCS equation with interaction $U(t)$ and a {\it real} order
parameter, also the chemical potential has to be adjusted according to Eq.~(\ref{eq:larm}).

Fig.~\ref{fig:4}a shows the time evolution of the phase for a given
adiabatic decay time $T=500$ when the chemical potential in $E_{BCS}$
is fixed (black curve).
Instead, when we transform to a time dependent chemical potential $\mu(t)$
according to Eq.~(\ref{eq:larm}) then the phase becomes time independent and equal to the initial value $\phi=0$ (red curve),
i.e. the order parameter is a real quantity during the time evolution.

We can also define a time-dependent chemical potential from the sign change
of $J_k^z$ motivated by the fact that this function develops into
the Fermi step function for large times (cf. Fig.~\ref{fig:3}b).
In the inset to Fig.~\ref{fig:3}b it becomes apparent that
the energy $\varepsilon_k$, for which $J_k^z=0$, shifts during the
dynamics of the system.
In fact, in the Larmor frame the transformed $z$-component of the pseudomagnetic field reads $b_z'=\varepsilon_k-\mu(t)-Un/2$
so that the solution for $b_z'=0$ is given by $J_k^{'x}=1/2$, $J_k^{'y}=0$, $J_k^{'z}=0$.

As can be seen from Fig.~\ref{fig:4}b, this leads to the same time dependent
$\mu(t)$ as obtained by transforming to the Larmor frame.
A further possibility is the computation of the time-dependent chemical
potential from the relation
\begin{displaymath}
  \mu(t)=\frac1N\frac{dE(n,t)}{dn}
\end{displaymath}
where $E(n,t)$ denotes the total {\it internal}  energy, not the grand canonical potential, $E_{BCS}(\mu,t)$.
$E(n,t)$ is obtained from  Eq.~(\ref{eq:ebdg})
setting $\mu_0=0$. Since $E(n,t)$ and $E(n+\delta n,t)$ are very close to
the equilibrium value in an adiabatic evolution, 
it is clear that ${dE(n,t)}/{dn}$ should also be close to the equilibrium chemical potential and the chemical potential obtained with the other criteria as shown in Fig.~\ref{fig:4}b.

\subsection{Restoring gauge invariance in BCS}\label{sec:ed}
While in the previous section we have considered the time
evolution of a spatially homogeneous system, we will now consider the
phase dynamics in case of a momentum dependent
electromagnetic field. In particular, we will be interested in the question
how, starting from an initially non-gauge invariant BCS state, gauge invariance is restored when we allow the phase to relax during the adiabatic time evolution.

\subsubsection{Setting up the problem}
In the presence of a static vector potential ${\bf A}({\qvec})$ the
total current should respond according to
\begin{equation}
  j_\mu(\qvec) =K_{\mu\nu}(\qvec) A_\nu(\qvec)
\end{equation}
where in the long-wave length limit
\begin{equation}\label{eq:kmunu}
  K_{\mu\nu}(\qvec)=\left\lbrack \delta_{\mu\nu}-\frac{q_\mu q_\nu}{q^2}\right\rbrack K(q^2)
\end{equation}
as a consequence of gauge invariance and charge conservation.\cite{Schafroth}

A superconductor and a conventional metal can be distinguished by
different limiting behavior of the transverse current response to an applied static vector
potential\cite{schrieff,scalapino93,and1,and2,gaugin} 
$\lim_{{\bf q}\to 0}{\bf j}({\bf q}) =\lim_{{\bf q}\to 0}D_s({\bf q}) {\bf A}({\bf q})$. 
 If the  limit ${\bf q} \to 0$ is taken for a momentum ${\bf q}\perp {\bf A}$, 
 then the superfluid stiffness $D_s$ is finite for a superconductor whereas it vanishes in the metal. However, for the BCS superconductor, defined from the
 variational wave-function  Eq. (\ref{eq:varbcs}) with a spatially
 uniform phase, one obtains
 the same result when the limit ${\bf q}\parallel {\bf A}$ is considered, whereas it should vanish in a gauge invariant approach [cf. Eq.~(\ref{eq:kmunu})], as  it does in the non-interacting metal.

 Without loss of generality we fix ${\bf A}(\qvec)\equiv A_x(\qvec) {\bf e}_x$
 and introduce the following definitions for the longitudinal and transverse responses
\begin{eqnarray}
  j_x(q_x\to 0,q_y=0)&=& D_l(q_x)A_x(q_x) \label{jqx}\\
  j_x(q_x=0, q_y\to 0) &=& D_t(q_y)A_x(q_y)\nonumber\\
  &=& D_s(1-\xi_t^2 q_y^2)A_x(q_y) \label{jqy} \,.
\end{eqnarray}
Here $D_s$ is the superfluid stiffness and we introduced $\xi_t$, the superconducting correlation length.

These investigations require the implementation of a momentum dependent
vector potential so that the evaluation of time dependent quantities
in terms of a density of states is not longer possible.
Our model system is a square lattice with
nearest-neighbor hopping and periodic boundary conditions for  $A_{x}$.

For evaluating the transverse response we therefore consider the coupling
\begin{equation}\label{eq:trans}
  \hat{T}_x=-t\sum_{ix,iy,\sigma}\left\lbrack c^\dagger_{ix,iy,\sigma}
  c_{ix+1,iy,\sigma}e^{-iA_{x}(iy)} + h.c. \right\rbrack
\end{equation}
with $A_{x}(iy)=A_0\cos(q_y i_y)$ and we leave translational invariance
along the $x$-direction.

The longitudinal coupling is given by
\begin{equation}\label{eq:long}
  \hat{T}_x=-t\sum_{ix,iy,\sigma}\left\lbrack c^\dagger_{ix,iy,\sigma}c_{ix+1,iy,\sigma}e^{-iA_{x}(ix)} + h.c. \right\rbrack
\end{equation}
with $A_{x}(ix)=A_0\cos(q_x i_x)$ and translational invariance is kept
along the $y$-direction.
In both cases the total current $j_x=-\partial\hat{T}_x/\partial A_{x}$
can be decomposed into a para- and diamagnetic part $j_x=j_{para}+j_{dia}$,
cf. Appendix \ref{a1}.

With these definitions the superconducting correlation length is, 
\begin{equation}\label{eq:xit}
  \xi_t^2=\frac{8t^4}{D_s}\frac{1}{N}\sum_k \sin^2(k_x)\sin^2(k_y)\frac{1}{E_k^3}\left(1-\frac{\varepsilon_k^2}{E_k^2}\right)
\end{equation}
as obtained from the current-current correlation function,
cf. e.g. Ref. \onlinecite{scalapino93}.
As will be shown explicitly below, the longitudinal vector potential can be eliminated from the problem so that physical observables do not depend on it.
Therefore,  in a gauge invariant approach,
\begin{equation}
  \label{eq:gi}
   D_l(q_x)=0\,.
\end{equation}

\begin{figure}[tb]
\includegraphics[width=8.5cm,clip=true]{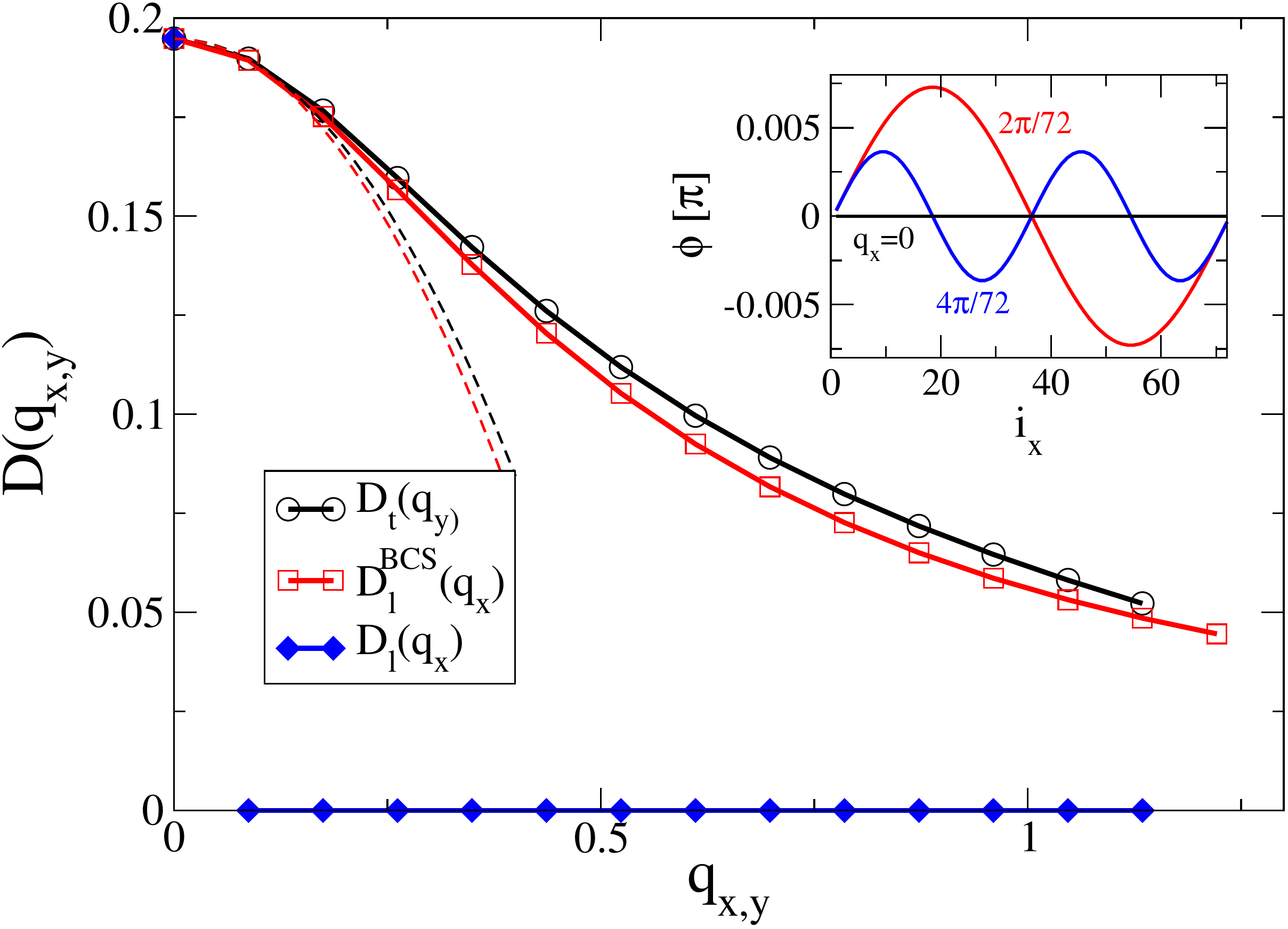}
\caption{Main panel: Static response to an external vector potential.
  Black: Transverse response to $A_x(q_y)$; Blue: (Gauge invariant) Longitudinal response to $A_x(q_x)$; Red: Longitudinal response in the BCS approximation.
  Dashed lines correspond to Eqs. (\ref{eq:xit}, \ref{eq:xil}) with
  $\xi_t=1.88$, $\xi_l=1.96$. Inset: Spatial variation of the order parameter phase for the (gauge invariant) longitudinal response. Parameters: $U/t=0.5$, $n=0.875$, $72\times 72$ lattice.}
\label{fig:5}                                                   
\end{figure}

As mentioned above, Eq.~(\ref{eq:gi}) is violated in the rigid-phase BCS approximation where 
\begin{equation}
  j^{BCS}_x(q_x\to 0,q_y=0)= D_s^0(1-\xi_l^2 q_x^2)A_x(q_x) \label{jqx2} \,.
\end{equation}
i.e. for $q_x\to 0$ it approaches the same limit than in the transverse case,
however, the correlation length is different and given by
\begin{equation}\label{eq:xil}
  \xi_l^2=\frac{8t^4}{D_s}\frac{1}{N}\sum_k \sin^4(k_x)\frac{1}{E_k^3}\left(1-\frac{\varepsilon_k^2}{E_k^2}\right) \,.
\end{equation}
It has been shown by Anderson \cite{and1,and2} and Rickayzen \cite{rick59}
that the longitudinal, in contrast to the transverse, response couples
to collective phase modes of the
order parameter and in this way gauge invariance can be restored.

In the normal metal the superfluid stiffness $D_s$ vanishes so that both, the
longitudinal and transverse response Eqs. (\ref{jqx},\ref{jqy}) vanish in this
case. It is therefore interesting to study the time dependent approach
of a BCS superconductor to the non-interacting limit and to follow the
time evolution of the longitudinal and transverse response.

\begin{figure}[tb]
\includegraphics[width=8.5cm,clip=true]{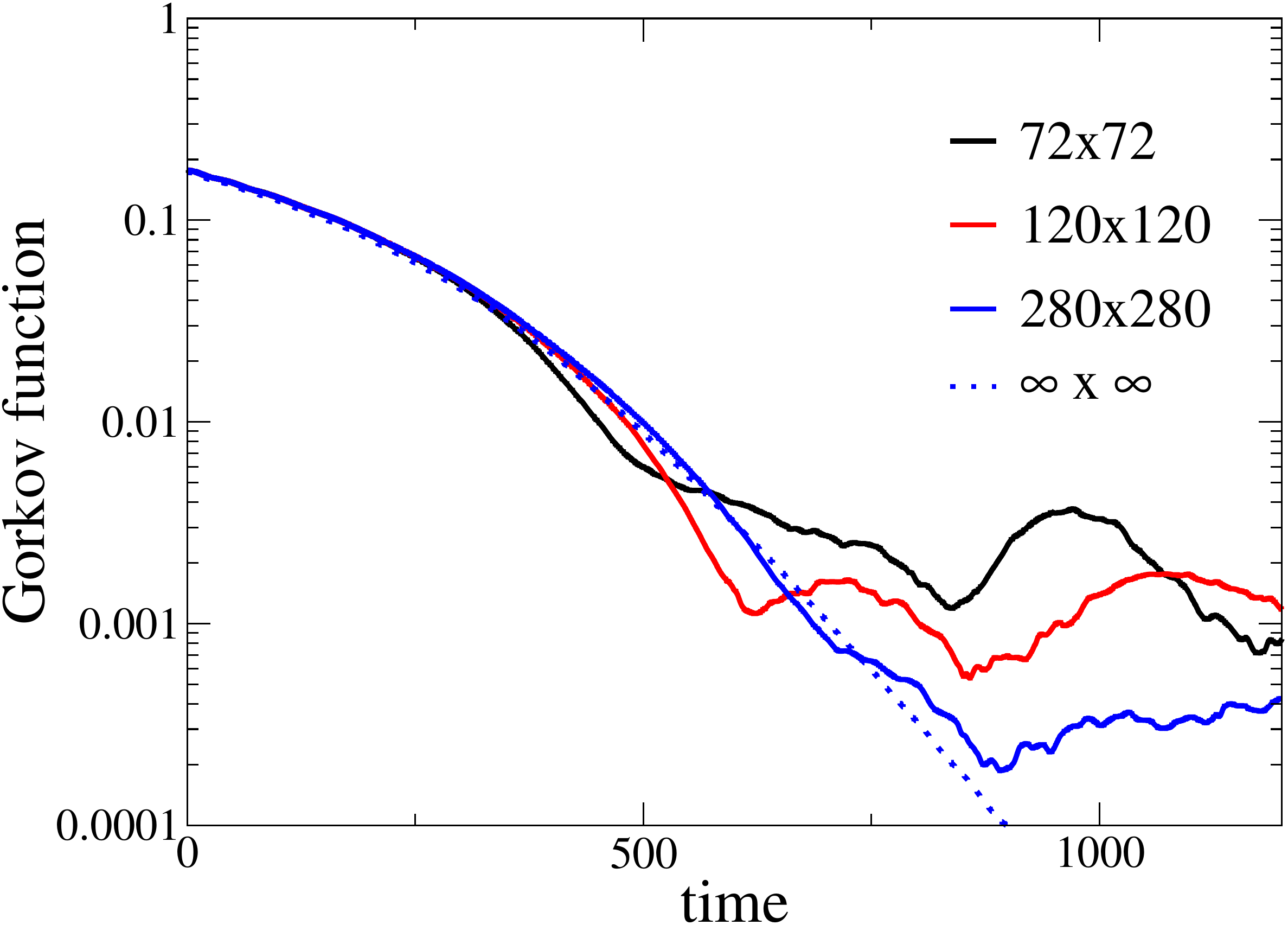}
\caption{Finite size effects for the adiabatic transition of the Gorkov function with $U(t)=U_0 e^{-t/T}$ and $T=500$. The dynamics for the infinite lattice
  blue dashed) has been computed with the 2-dim DOS. 
Parameters: $U_0/t=0.5$, $n=0.875$.}
\label{fig:6}                                                   
\end{figure}

Figure~\ref{fig:5} compares the transverse and
longitudinal equilibrium response for a finite system and demonstrates the agreement with the expansion
at small momenta given by Eqs. (\ref{eq:xit},\ref{eq:xil}). 

\subsubsection{Irrelevance of a longitudinal vector potential}
The problem of the superconductor coupled to a longitudinal vector potential
can be mapped to a superconductor without such coupling. Thus, the longitudinal vector potential can be ``gauged away''  and is physically irrelevant. 
Within our definition of the coupling Eq. (\ref{eq:long}), the gauge
transformation to new operators $c_{ix,iy,\sigma}'$ reads, 
$$
c_{ix,iy,\sigma}'=c_{ix,iy,\sigma}\exp\left[-i\sum_{i_x'=0}^{i_x-1} A_{x}(i_x')\right].   
$$
where $i_x=0,..,L-1$ and $A_x(L)=A_x(0)$.
This relation implies that
$$
c_{L,iy,\sigma}'=c_{L,iy,\sigma}\exp\left[-i\sum_{i_x'=0}^{L-1} A_{x}(i_x')\right]=c'_{0,iy,\sigma}.   
$$
which allows to close the boundary conditions without any effect of the longitudinal field. In terms of the new operators the kinetic energy reads
like Eq.~(\ref{eq:long}) but without any vector potential.
Since for local observables the phase factors are irrelevant, 
it follows that any physical observable can not depend on the longitudinal
field.

We can now use the inverse transformation to understand the equilibrium
solution in the original variables. Once the vector potential has been
eliminated, we can assume an equilibrium Gorkov function with
spatially uniform phase $f'={const.}$
Transforming back to the old operators one obtains
a Gorkov function, 
\begin{equation}\label{eq:ftrans}
f(i_x)=f'\exp \left[2i\sum_{i_x'=0}^{i_x-1} A_{x}(i_x')\right]
\end{equation}
so that gauge invariance completely determines the phase of the order
parameter in the original variables. 
Since  $A_{x}(i_x')\propto \cos(q\, i_x)$ for small $q$ we can use the continuum limit to show that the phase of the Gorkov function should behave as $\phi\propto  \sin(q \,i_x)/q$  
(cf. inset to Fig.~\ref{fig:5}). Clearly, this is in contrast with the rigid-phase BCS approximation where $\phi\equiv 0$  
yields the incorrect result $D_l(q_x\to 0,q_y=0)=D_t(q_x=0,q_y\to 0)$
which is equivalent to the (negative) kinetic energy
along the direction of the applied vector potential.

Due to the above mapping, the adiabatic evolution of a system with a longitudinal vector potential is equivalent to the corresponding dynamics of a uniform
system. The phase of the order parameter is related to the uniform case
by Eq. (\ref{eq:ftrans}), so the spatial modulation does not depend on time.
It is therefore more interesting to study the time evolution
when the initial state is prepared in the non-gauge invariant BCS
state Eq. (\ref{eq:varbcs}) with a uniform phase,  and gauge invariance
has to be restored during the time evolution towards the non-interacting
system.
In the transverse case a similar gauge transformation cannot be done as
$\sum_{i_x'=1}^{L} A_{x}(i_y)\ne 0$. The adiabatic dynamics of these two
cases is studied below.

\subsubsection{Finite size effects}

The necessity of considering finite lattices implies finite size effects in
the time evolution which
are shown in Fig.~\ref{fig:6} for the dynamics of the Gorkov function $|f|$. At
long times this function does not approach $|f|=0$ as expected for the
infinite system but one observes oscillations around finite values due
to incomplete dephasing. Therefore results below are only strictly
valid for a limited time interval, nevertheless we can draw some conclusions
about the general behavior at large times.

\begin{figure}[h!]
\includegraphics[width=8.5cm,clip=true]{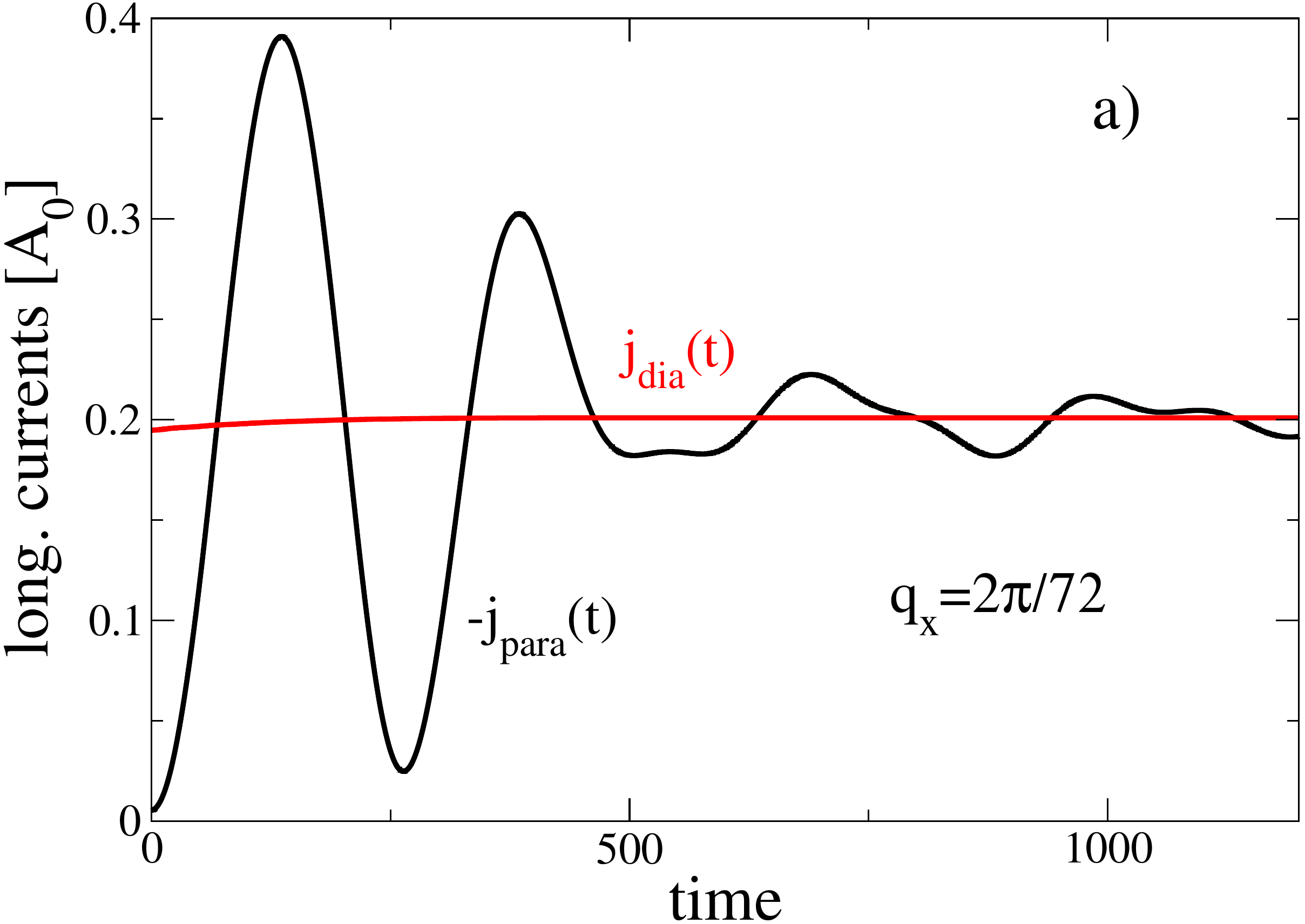}
\includegraphics[width=8.5cm,clip=true]{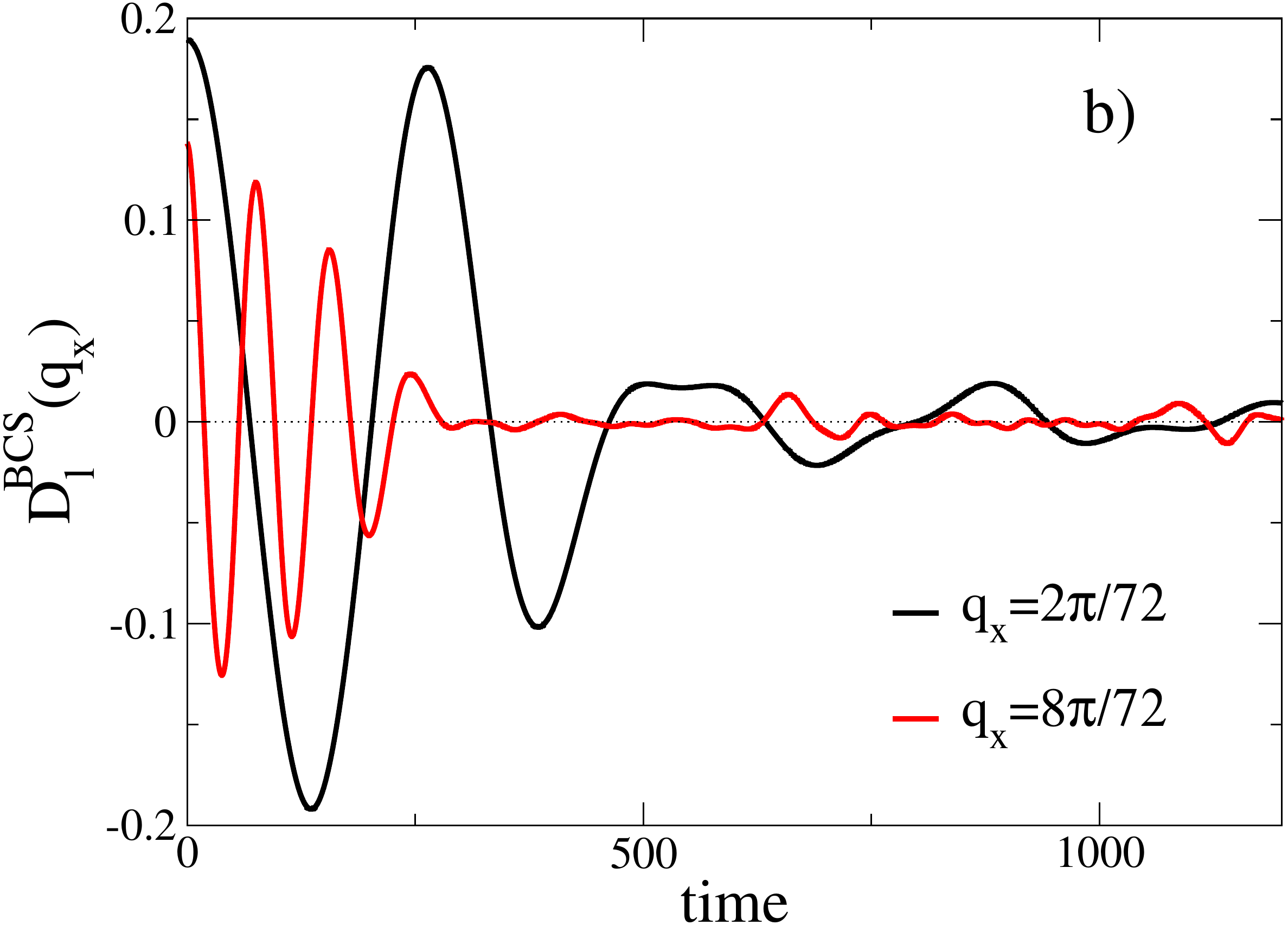}
\caption{Time dependence of the longitudinal response when the initial
  state is computed in the BCS approximation. a) Dia- (red) and paramagnetic (black) currents
  for $q_x=2\pi/72$. b) Total response $D_l=(j_{para}+j_{dia})/A_0$ for
  for $q_x=2\pi/72$ (black) and $q_x=8\pi/72$ (red).
Parameters: $U_0/t=0.5$, $n=0.875$. $72\times 72$ lattice.}
\label{fig:7}                                                   
\end{figure}

\subsubsection{Restoring gauge invariance with a longitudinal field}
For an exponential  decay of the interaction $U=U_0 e^{-t/T}$ with
$T=500$, Fig.~\ref{fig:7} reports the longitudinal response when the initial
state is prepared in the non-gauge invariant BCS state Eq. (\ref{eq:varbcs}),
i.e. the order parameter
phase is set to $\phi=const. \equiv 0$. In the time evolution instead,
we allow the phase to relax. Therefore, even in the presence of $A(i_x)$, the
initial paramagnetic  current vanishes, $j_{para}(t=0)=0$,
whereas the diamagnetic
current corresponds to the kinetic energy $-\langle t_x\rangle$ of the
initial BCS solution. For large times the latter approaches $-\langle t^0_x\rangle$ of
the non-interacting system whereas the paramagnetic current undergoes
decaying low frequency oscillations. The total response (cf. Fig.~\ref{fig:7}b)
therefore decays from the initial $D_l(q_x,t=0)=-\langle t_x\rangle(t=0)$
to $D_l(q_x,t\to \infty)\approx 0$ thus restoring gauge invariance at large
times.

\begin{figure}[tb]
\includegraphics[width=8.5cm,clip=true]{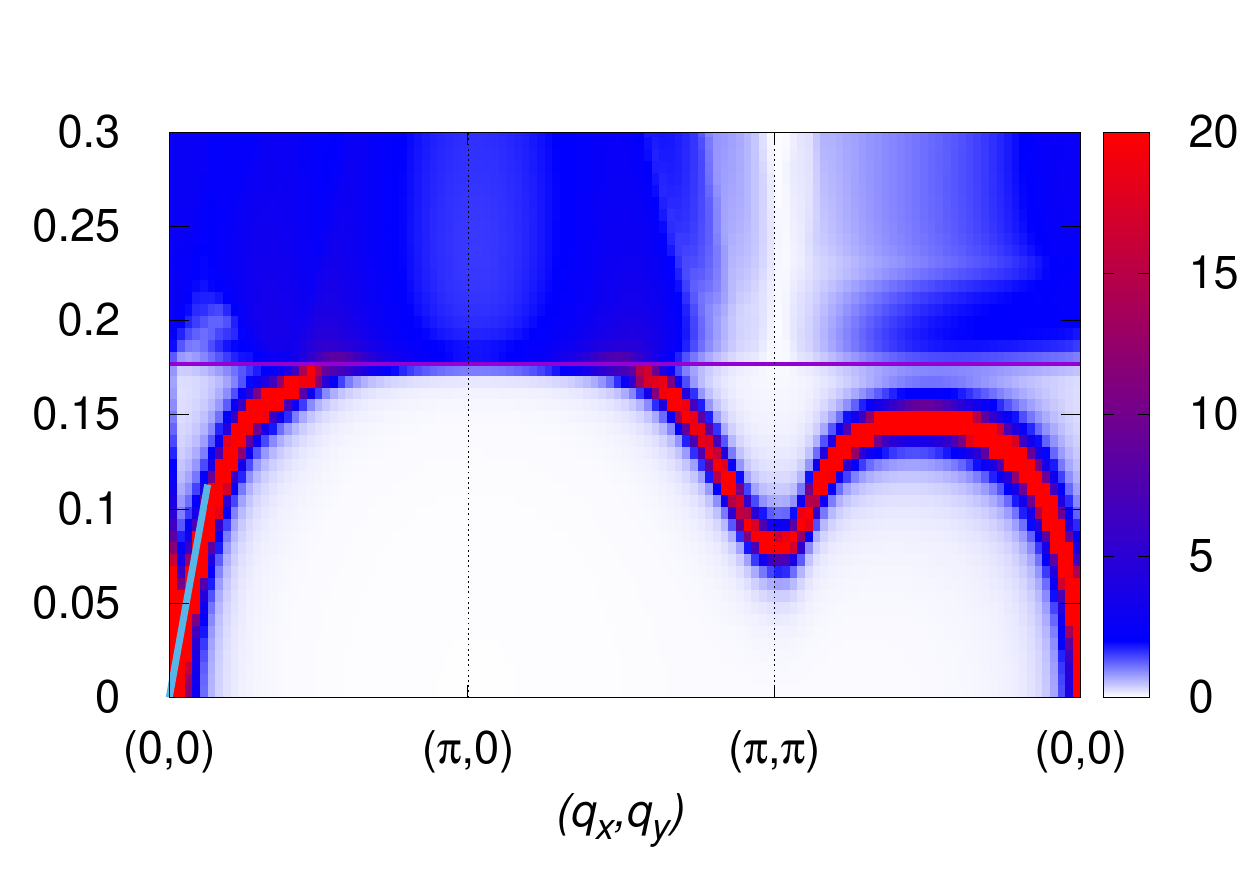}
\caption{Phase mode along a selected momentum cut in the Brillouin zone.
  The linear approximation at small momenta is $\omega_q=0.25 q$
  and fits to the observed phase excitations in the current shown in
Fig.~\ref{fig:7}. Parameters: $U/t=0.5$, $n=0.875$.}
\label{fig:8}                                                   
\end{figure}

 The frequency of the initial oscillations in  $j_{para}$ linearly increases
with $q_x$ and can be attributed to phase modes. In Fig.~\ref{fig:8} we
show the dispersion of phase excitations across the Brillouin zone
for the same parameters, as obtained from an RPA calculation on top of the
initial BCS state, and it turns out that the oscillations appearing
in Fig.~\ref{fig:7} fit to the sound mode $\omega\approx 0.113 q_x$ which is
shown as the linear fit at small momenta in Fig.~\ref{fig:8}.

\begin{figure}[h!]
\includegraphics[width=8.5cm,clip=true]{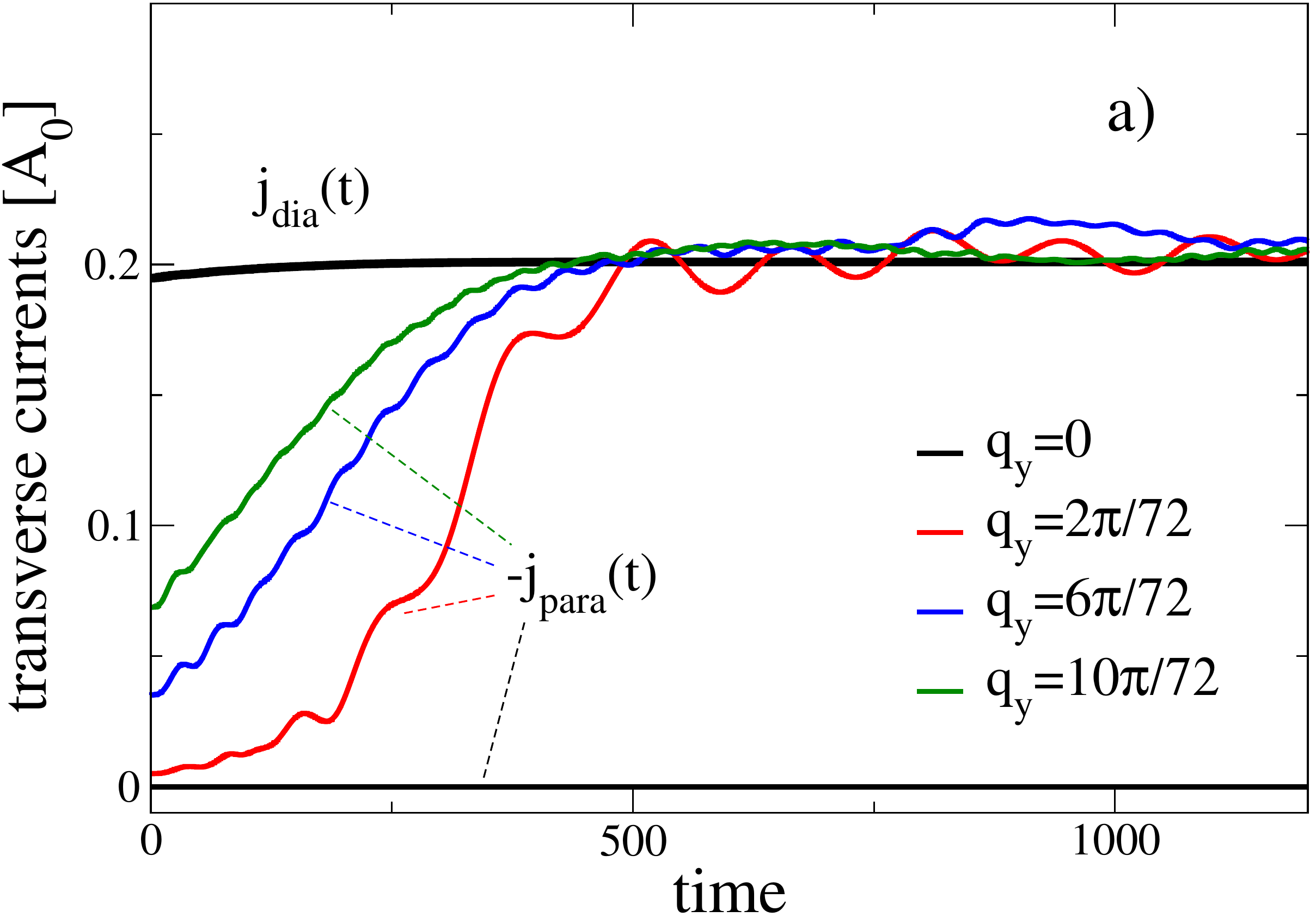}
\includegraphics[width=8.5cm,clip=true]{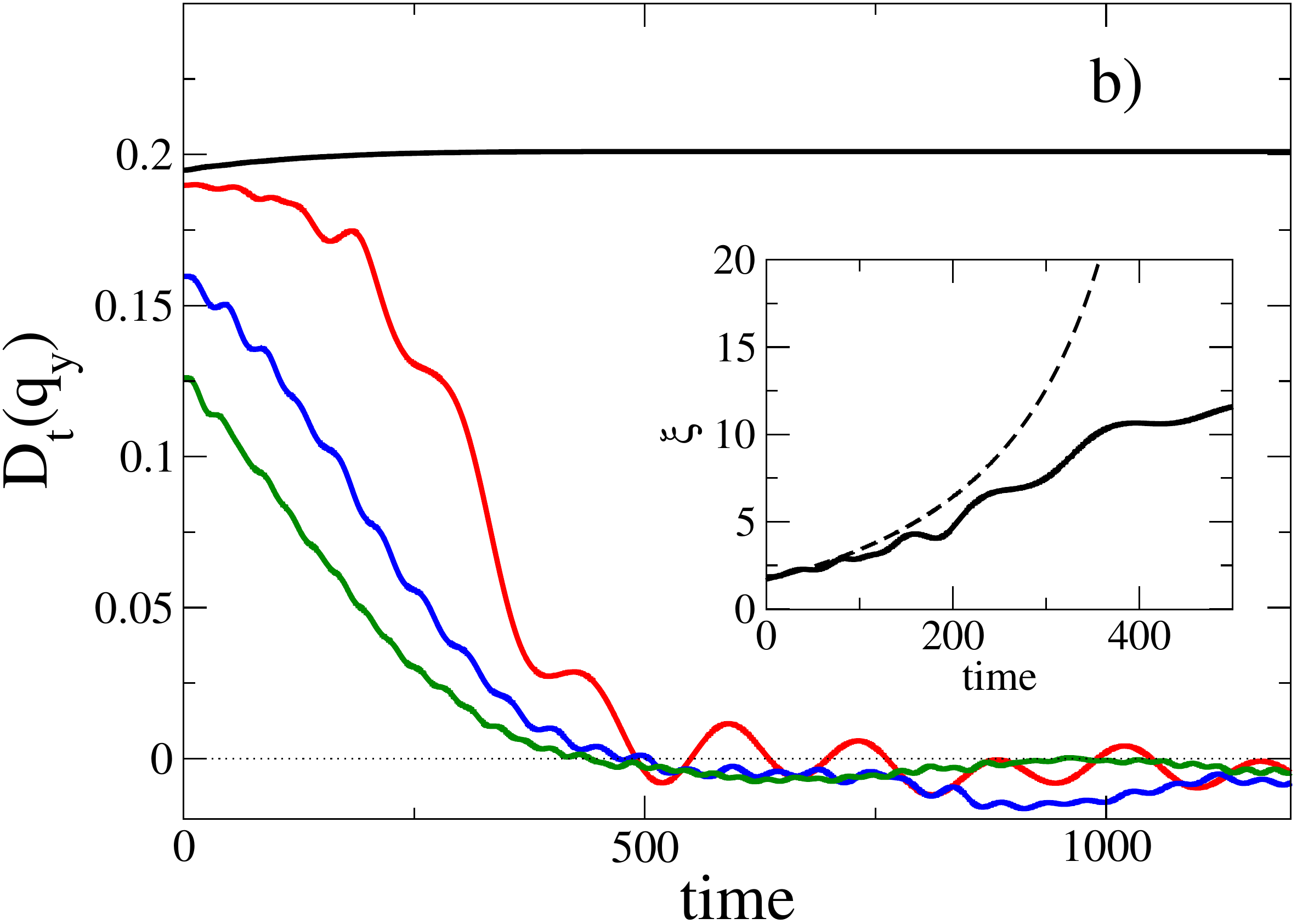}
\caption{ 
Time dependence of the transverse response. 
  a) Dia- and paramagnetic currents for different
  momenta $q_y$. b) Total response $D_t(q_y,t)=(j_{para}+j_{dia})/A_0$.
  The inset compares the time dependent correlation length (solid)
  with the linear response expression Eq.~(\ref{eq:xit}). 
Parameters: $U_0/t=0.5$, $n=0.875$. $72\times 72$ lattice.}
\label{fig:9}                                                   
\end{figure}  

\subsubsection{Time-dependent transverse response}
The transverse response is analyzed in Fig.~\ref{fig:9}. Because of London rigidity, the phase is constant in the presence of the transverse vector potential. We find that also in the non-equilibrium state the phase does not couple to the response, i.e. also for an applied vector potential $A_x(q_y)$ the
order parameter phase stays spatially homogeneous. Thus, the dynamics occurs purely on the BCS level and, as a consequence, the time-dependent
response at zero transverse momentum $D_s^0(q_y=0)$ is simply
given by the time-dependent kinetic energy $-\langle t_x\rangle$ along the direction of the applied vector potential.
This is shown in Fig.~\ref{fig:9}a which reports 
the time evolution of the paramagnetic current for different
momenta $q_y$ of the applied vector potential. 

Since the relation between diamagnetic current and vector potential is local
(i.e. ${\bf j}_{dia}(ix,iy)\propto {\bf A}(ix,iy)$ the diamagnetic
contribution (black curve) is independent of momentum.

At $q_y=0$ the paramagnetic current vanishes whereas $j_{dia}/A_0$ evolves
from the kinetic energy of the BCS superconductor  to the
kinetic energy of the non-interacting metal.
For finite $q_y$ and in the long-time limit $-j_{para}$ approaches the
diamagnetic current
so that in this limit the total finite momentum response tends to zero,
cf. panel (b).
Thus, the transition from a superconducting to normal state via the
adiabatic transition is completed when $D_t(q_y)=0$ for the smallest 
momentum $q_y$ so that $\lim_{q_y\to 0}D_t(q_y)=D_s^0=0$, i.e. zero
superfluid stiffness. Note that in our case, due to finite size effects,
$D_t(q_y,t\to \infty)$ approaches a small finite value only.

In fact, the transition should be accompanied by a diverging
correlation length which in a finite lattice is always limited
by the system size, i.e. $\xi_{max}=1/q_y^{min}$. This is shown in the
inset to panel (b) which compares $\xi$ with the corresponding linear
response result Eq.~(\ref{eq:xit}). For our finite lattice $\xi$ tends
to saturate at $\xi=72/(2\pi) \approx 12$ (solid line)
whereas the thermodynamic result (dashed) shows the expected divergence.

\section{Discussion and Conclusions}
In this paper we have performed a comprehensive analysis
of the phase dynamics when a BCS superconductor evolves
into a non-interacting Fermi liquid by means
of an exponentially decaying pairing interaction $U(t)$
with time scale $T$.
Adiabaticity is obeyed when at each instant of time the
order parameter obeys the corresponding BCS equation for
the instantaneous interaction parameter $U(t)$.
We have shown that for times $t \lesssim T$ this condition is fulfilled,
which agrees with the condition that in this regime the rate of change of the
gap is smaller than the square of the gap.~\cite{polkov11} Thus, in order to
preserve adiabaticity the 
change in $U$ should be slower than the closing of the gap and  a Fermi
liquid with $\Delta=0$ can only
be reached in the limit $T\to \infty$. 

One can make the evolution slow enough so that at long times
$t\gg T$, where the gap becomes negligibly small,
the momentum distribution appears as a step within a given resolution.
However, in the thermodynamic limit, memory of the superconducting origin of this ``Fermi liquid'' is still preserved in the orientation of Anderson pseudospins
close to the Fermi level.  This was demonstrated by showing that  one can fully recover the
original BCS state by inverting the dynamics at a very long time.
In this context, we have discussed the analogy between the reversed dynamics and  spin/photon echo experiments. In principle, the inversion of time evolution
can be experimentally investigated in
ultracold quantum gases where extensive control of interactions is
possible.~\cite{lv20}

The memory of the superconducting state can be erased if one inverts the
order of limits, i.e. by first considering $t\to \infty$ for a finite system
then taking the thermodynamic limit.
In this case one reaches a perfect metal at $T=0$ in which
all levels are either empty or occupied, i.e. for each state
the Anderson pseudospins are oriented along $\pm z$.

We also studied the evolution of the phase, which in our system is not
constrained to be constant by particle hole-symmetry. 
The requirement, that at each instant of time the solution can be
described by the BCS equation with a real valued order parameter requires
the implementation of a time-dependent chemical potential $\mu(t)$.
We have shown that formally $\mu(t)$
can be obtained from the time-derivative of the phase $\dot{\phi}$
since phase and charge are conjugate variables in the problem.
The same $\mu(t)$ results
from the sign change of the pseudospin variable $J_k^z$ which in the
adiabatic limit evolves into the zero temperature Fermi distribution.
Moreover, it turned out that the same time-dependent chemical potential
obeys the thermodynamic relation $\mu(t)=\partial E/\partial n$.

Finally, we have explicitely shown how gauge invariance is restored when
the BCS superconductor with broken $U(1)$ symmetry, evolves into
the Fermi liquid. The longitudinal response to an applied vector potential
is corrected by the coupling to phase fluctuations which guarantee the
vanishing of the corresponding current at smallest momenta and large times.
For the transverse coupling it turned out that the phase does not
couple, so that the corresponding physics in the time dependent situation
is the same than in the stationary case. It has has been reported \cite{yus06}
that in a quench situation where
the order parameter drops to zero the superfluid stiffness reduces
to half of $-\langle t_x\rangle$, however, this result seems to be incompatible
with the present findings. This may be due to the fact that the response in
Ref. \onlinecite{yus06} has only be considered for zero momentum, whereas, as detailed in Sec. \ref{sec:ed}, the superfluid stiffness has to be formally
evaluated from the proper zero momentum limit.

\acknowledgements
G.S. acknowledges financial support from the Deutsche Forschungsgemeinschaft under SE 806/19-1. J.L. acknowledges financial support from Italian Ministry for University and Research through PRIN Project No. 2017Z8TS5B and Regione Lazio (L. R. 13/08) through project SIMAP.

\appendix
\section{Evaluation of the transverse and longitudinal response}
\label{a1}

\subsection{Transverse response}
For the considered coupling to a transverse vector potential in
Eq.~(\ref{eq:trans}) one can perform a Fourier transformation along the $x$-direction to the resulting hamiltonian $H=\sum_{kx} H(k_x)$ with
\begin{eqnarray}
  H(k_x)&=& -2t\sum_{iy}\left\lbrack \cos(k_x-A_x(i_y))c_{kx,\uparrow}^\dagger(i_y) c_{kx,\uparrow}(i_y)\right. \nonumber \\
  &-&\left. \cos(k_x+A_x(i_y))c_{-kx,\downarrow}(i_y) c^\dagger_{-kx,\downarrow}(i_y) \right\rbrack\nonumber \\
  &-&t\sum_{iy}\left\lbrack c^\dagger_{kx,\uparrow}(i_y+1)c_{kx,\uparrow}(i_y)\right. \nonumber \\
  &-&\left. c_{-kx,\downarrow}(i_y+1)c^\dagger_{-kx,\downarrow}(i_y) + h.c. \right\rbrack \nonumber \\
  &+&\sum_{iy}\left\lbrack \Delta c^\dagger_{kx,\uparrow}(i_y)c^\dagger_{-kx,\downarrow}(iy) +h.c. \right\rbrack \nonumber \\
  &-& |U(t)|\sum_{iy}\frac{n_{iy}}{2}\left\lbrack c^\dagger_{kx,\uparrow}(i_y)c_{kx,\uparrow}(i_y)\right. \nonumber \\
  &-&\left. c_{kx,\downarrow}(i_y)c^\dagger_{kx,\downarrow}(i_y) \right\rbrack \label{eq:a1}
  \end{eqnarray}

For each momentum $k_x$ the hamiltonian can be diagonalized
by a Bogoljubov de-Gennes (BdG) transformation
\begin{equation}
  c_{kx,\sigma}(iy)=\sum_p\left\lbrack u_{iy,kx}(p)\gamma_{p,\sigma}(k_x)-\sigma v_{iy,kx}(p)^* \gamma_{p,-\sigma}^\dagger(k_x)\right\rbrack
\end{equation}
which for each momentum allows to construct the density matrix $\underline{\underline{R}}(k_x)$. For a $N\times N$ lattice it has dimension $2N\times 2N$
and obeys the equation of motion Eq.~(\ref{eq:densmat}).

In order to define dia- and paramagnetic currents we expand Eq. (\ref{eq:a1})
in powers of the vector potential
\begin{eqnarray*}
  H(k_x)&=&H(k_x,A_x=0)\\
  &-&\sum_{iy} \left\lbrack j^x_{para}(k_x,i_y)A_x(i_y)
  +\frac{1}{2}j^x_{dia}(k_x,i_y)A^2_x(i_y) \right\rbrack
\end{eqnarray*}
with
\begin{eqnarray*}
  j^x_{para}(k_x,i_y)&=&2t\sum_{\sigma}\sin(k_x)\langle c^\dagger_{kx,\sigma}(i_y)c_{kx,\sigma}(i_y)\rangle \\
  j^x_{dia}(k_x,i_y)&=&-2t\sum_{\sigma}\cos(k_x)\langle c^\dagger_{kx,\sigma}(i_y)c_{kx,\sigma}(iy)\rangle 
\end{eqnarray*}  
and the total current is given by
\begin{eqnarray}
  j_{tot}^x(k_x,i_y)&=&-\frac{\delta H(k_x)}{\delta{A_x(i_y}} \\
  &=& j^x_{para}(k_x,i_y)+j^x_{dia}(k_x,i_y)A_x(i_y) \nonumber \,.
\end{eqnarray}
The currents shown in Fig. \ref{fig:9} correspond to the
Fourier transformed quantities at $\pm q_c$ and thus are given by
\begin{eqnarray*}
  j^x_{para}(q_y)&=&\frac{2t}{N}\sum_{\sigma,iy,kx}\sin(k_x)\langle c^\dagger_{kx,\sigma}(i_y)c_{kx,\sigma}(i_y)\rangle e^{i q_y i_y} \\
  j^x_{dia}(q_y)&=&-\frac{2t A_0}{N}\sum_{\sigma,iy,kx}\cos(k_x)\langle c^\dagger_{kx,\sigma}(i_y)c_{kx,\sigma}(i_y)\rangle 
\end{eqnarray*}

  \subsection{Longitudinal response}
For the longitudinal coupling Eq.~(\ref{eq:long})
the Fourier transformation is performed along the $y$-direction
and the resulting hamiltonian $H=\sum_{ky} H(k_y)$ reads
\begin{eqnarray}
  H(k_y)&=& -t\sum_{ix}\left\lbrack e^{-iA_x(ix)}c_{ky,\uparrow}^\dagger(i_x) c_{ky,\uparrow}(i_x+1)\right. \nonumber \\
  &-&\left. e^{iA_x(ix) }c_{-ky,\downarrow}(i_x) c^\dagger_{ky,\downarrow}(i_x+1) + h.c. \right\rbrack\nonumber \\
  &-& 2t\sum_{ix}\cos(k_y) \left\lbrack c^\dagger_{ky,\uparrow}(i_x)
  c_{ky,\uparrow}(i_x)\right. \nonumber \\
  &-&\left. c_{-ky,\downarrow}(ix)c^\dagger_{-ky,\downarrow}(ix)  \right\rbrack \nonumber \\
  &+&\sum_{iy}\left\lbrack \Delta(i_x) c^\dagger_{ky,\uparrow}(i_x)c^\dagger_{-ky,\downarrow}(ix) +h.c. \right\rbrack \nonumber \\
  &-& |U(t)|\sum_{ix}\frac{n_{ix}}{2}\left\lbrack c^\dagger_{ky,\uparrow}(i_x)c_{ky,\uparrow}(i_x)\right. \nonumber \\
  &-&\left. c_{ky,\downarrow}(ix)c^\dagger_{ky,\downarrow}(i_x) \right\rbrack
  \end{eqnarray}
which can be diagonalized by a similar BdG transformation than in the
transverse case.

For the longitudinal case we obtain the dia- and paramgnetic response from
\begin{eqnarray*}
  j^x_{dia}(q_x)&=&-\frac{t A_0}{N}\sum_{ky,ix,\sigma}\left(\langle c_{ky,\uparrow}^\dagger(ix) c_{ky,\uparrow}(ix+1)\rangle +h.c.\right)\\
  j^x_{para}(q_x)&=&\frac{it}{N}\sum_{ky,ix,\sigma}\left(\langle c_{ky,\uparrow}^\dagger(ix) c_{ky,\uparrow}(ix+1)\rangle - h.c.\right) e^{i q_x i_x} \,.
\end{eqnarray*}

\section{Dynamics for an increasing interaction}
\label{a2}

\begin{figure}[hhh]
  \includegraphics[width=8.5cm,clip=true]{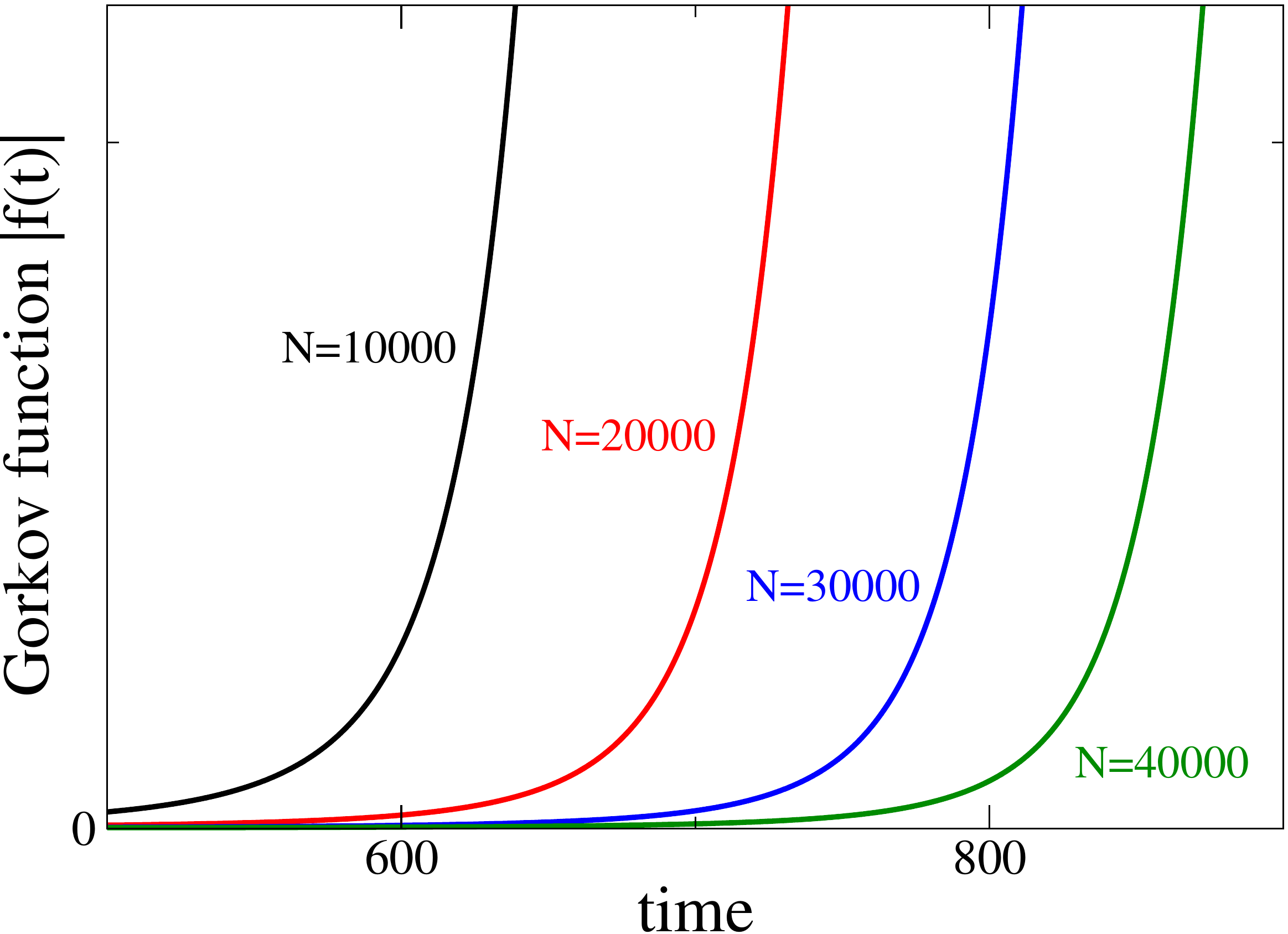}
\caption{Time dependence of the Gorkov function with the
  initial density matrix given by Eq.~(\ref{eq:dens0}),(\ref{eq:dens1})
  and a time dependent interaction $U(t)=U_0 (\exp(t/T)-1)$. 
  Parameters: $\varphi=0$, $U_0=0.5$, $T=1000$.} 
\label{fig:2a}                                                   
\end{figure}  

Fig.~\ref{fig:2a} shows the time evolution of the Gorkov function
$f(t)$ for an exponentially increasing interaction and an initial
density matrix given by Eqs.~(\ref{eq:dens0},\ref{eq:dens1}). Note that this does
{\it not} correspond to the inverse time evolution of the dynamics shown in
Fig.~\ref{fig:1} since in this case the state Eq.~(\ref{eq:dens0})
would only be reached for $t\to \infty$.
Since one has a single 'nucleation state' for superconductivity at
the chemical potential the time evolution of $f(t)$ is clearly dependent
on the number of energy levels $N$ so that in the limit of $N\to \infty$
the onset of exponential growth of $f(t)$ would be shifted to $t\to\infty$,
i.e. one would stay in the Fermi liquid state for all times.


\begin{thebibliography}{cc}
\bibitem{born28}M. Born and V. A. Fock, Z. Phys. A {\bf 51}, 165180 (1928).
\bibitem{gml51} M. Gell-Mann and F. Low, Phys. Rev. {\bf 84}, 350 (1951).
\bibitem{giorg08} S. Giorgini, L. P. Pitaevskii, and S. Stringari, Rev. Mod. Phys. {\bf 80}, 1215 (2008).
\bibitem{toermae18} P. T\"orm\"a, Phys. Scr. {\bf 91}, 043006 (2016).
\bibitem{polkov05} A. Polkovnikov, Phys. Rev. B {\bf 72}, 161201 (2005).
\bibitem{dziarmaga05} J. Dziarmaga, Phys. Rev. Lett. {\bf 95}, 245701 (2005).
\bibitem{zurek05} W. H. Zurek, U. Dorner, and P. Zoller, Phys. Rev. Lett. {\bf 95}, 105701 (2005).
\bibitem{kuhn1} T. Papenkort, V. M. Axt, and T. Kuhn, Phys. Rev. B {\bf 76}, 224522 (2007).  
\bibitem{kuhn2} T. Papenkort, T. Kuhn, and V. M. Axt, Phys. Rev. B {\bf 78}, 132505 (2008).
\bibitem{mazza12} G. Mazza and M. Fabrizio, Phys. Rev. B {\bf 86}, 184303 (2012).
\bibitem{manske14} H. Krull, D. Manske, G. S. Uhrig, and A. P. Schnyder, Phys. Rev. B {\bf 90}, 014515 2014).
\bibitem{Bunemann2017}
J. B{\"{u}}nemann and G. Seibold, Phys. Rev. B {\bf 96},  245139  (2017).
\bibitem{mazza17} G. Mazza, Phys. Rev. B {\bf 96}, 205110 (2017).
\bibitem{Collado2018} H. P. O. Collado, J. Lorenzana, G. Usaj, and C. A. Balseiro, Phys. Rev. B {\bf 98}, 214519 (2018).
\bibitem{Collado2019}
H.~P. {Ojeda Collado}, G. Usaj, J. Lorenzana, and C.~A. Balseiro, Phys. Rev. B
  {\bf 99},  174509  (2019).
\bibitem{Collado2020}
  H.~P. {Ojeda Collado}, G. Usaj, J. Lorenzana, and C.~A. Balseiro, Phys. Rev. B
   {\bf 101},  054502  (2020).
\bibitem{benfatto19} M. Udina, T. Cea, and L. Benfatto, Phys. Rev. B {\bf 100}, 165131 (2019).
\bibitem{goetz20} G. Seibold and J. Lorenzana, Phys. Rev. B {\bf 102}, 2469 (2020).
\bibitem{and2} P. W. Anderson, Phys. Rev. {\bf 112}, 1900 (1958).  
\bibitem{bara04} R. A. Barankov, L. S. Levitov, and B. Z. Spivak,  Phys. Rev. Lett. {\bf 93}, 160401 (2004).
\bibitem{yus05}  E. A. Yuzbashyan, B. L. Altshuler, V. B. Kuznetsov, and V. Z. Enolskii, J. Phys. A: Math. Gen. {\bf 38}, 7831 (2005).
\bibitem{bara06} R. A. Barankov and L. S. Levitov, Phys. Rev. Lett. {\bf 96},
  230403 (2006).
\bibitem{alt06} E. A. Yuzbashyan, O. Tsyplyatyev, and B. Altshuler, Phys. Rev. Lett. {\bf 96}, 097005 (2006).
\bibitem{yus06} E. A. Yuzbashyan and M. Dzero, Phys. Rev. Lett. {\bf 96}, 230404 (2006).
\bibitem{hahn} E. L. Hahn, Phys. Rev. {\bf 80}, 580 (1950).
\bibitem{bcs} J. Bardeen, L. N. Cooper, and J. R. Schrieffer, Phys. Rev. {\bf 108}, 1175 (1957).  
  \bibitem{Blaizot1986}
J.~P. Blaizot and G. Ripka, {\em {Quantum Theory of Finite Systems}} (The MIT
  Press, Cambridge, Massachusetts, 1986), pp.\ 1--657.
\bibitem{and1} P. W. Anderson, Phys. Rev. {\bf 110}, 827 (1958).
\bibitem{gaugin} N. N.
Bogoliubov,  V. V. Tolmachov, and D. V. Širkov, Fortschr. Phys. {\bf 6}, 605–682 (1958);
Y. Nambu, Phys. Rev. {\bf 117}, 648–663 (1960).
\bibitem{Scaramazza2019}
J.~A. Scaramazza, P. Smacchia, and E.~A. Yuzbashyan, Phys. Rev. B {\bf 99},
054520  (2019).
\bibitem{OjedaCollado2021}
H.~P. {Ojeda Collado}, G. Usaj, C.~A. Balseiro, D.~H. Zanette, and J.
  Lorenzana, Emergent parametric resonances and time-crystal phases in driven BCS systems, arXiv:2107.09683.
\bibitem{polkov11} A. Polkovnikov, K. Sengupta, A. Silva, and M. Vengalattore, Rev. Mod. Phys. {\bf 83}, 863 (2011).  
\bibitem{yuz04}  E. A. Yuzbashyan and M. Dzero, Phys. Rev. Lett. {\bf 96}, 230404 (2004).
\bibitem{Schafroth} M. R. Schafroth, Helv. Phys. Acta {\bf 24}, 645 (1951).
\bibitem{schrieff} J. R. Schrieffer, {\it Theory of Superconductivity}, Westview Press (1964).
\bibitem{scalapino93} D. J. Scalapino, S. R. White, and S. Zhang, Phys. Rev. B {\bf 47}, 7995 (1993).
\bibitem{rick59} G. Rickayzen, Phys. Rev. {\bf 115}, 795 (1959).
\bibitem{lv20} C. Lv, R. Zhang, and Q. Zhou, Phys. Rev. Lett. {\bf 125}, 253002 (2020).

\end{thebibliography}

\end{document}